\documentclass[12pt,reqno]{amsart}

\usepackage[foot]{amsaddr}
\usepackage{booktabs}

\usepackage{amsthm}
\usepackage{tabularx}

\usepackage{mathtools}

\usepackage{amssymb}
\usepackage{amsmath}
\usepackage{amsfonts}
\usepackage{mathrsfs}
\usepackage{nicefrac}
\usepackage[dvipdfmx]{graphicx} 
\usepackage{threeparttable}
\usepackage[referable]{threeparttablex}
\usepackage{subcaption}
\usepackage{bmpsize}
\usepackage{color}
\usepackage[onehalfspacing]{setspace}
\usepackage{caption}
\usepackage{natbib}
\usepackage[shortlabels]{enumitem}
\usepackage[utf8]{inputenc}
\usepackage{charter}
\usepackage[colorlinks=true,citecolor=blue,urlcolor=blue,pdfpagemode=UseNone,pdfstartview=FitH]{hyperref}
\usepackage{apptools}
\usepackage{color}
\usepackage{subcaption}
\usepackage{bbm}

\makeatletter
\def\section{\@startsection{section}{1}
	\z@{1.0\linespacing\@plus\linespacing}{.8\linespacing}{\Large}}

\def\subsection{\@startsection{subsection}{2}
	\z@{.8\linespacing\@plus.7\linespacing}{.7\linespacing}{\large}}

\def\subsubsection{\@startsection{subsubsection}{3}
	\z@{.5\linespacing\@plus.7\linespacing}{-.5em}{\normalfont\bfseries}}
\makeatother

\numberwithin{equation}{section}

\theoremstyle{definition}

\theoremstyle{definition}

\theoremstyle{definition}

\theoremstyle{definition}

\allowdisplaybreaks
\usepackage{setspace}

\usepackage{multirow}

\expandafter\let\expandafter\oldproof\csname\string\proof\endcsname
\let\oldendproof\endproof
\renewenvironment{proof}[1][\proofname]{%
	\oldproof[\normalfont \bfseries  #1]%
}{\oldendproof}

\title{Rethinking Cost-Sharing Policies: Enhancing Chronic Disease Management for Disadvantaged Populations}
\thanks{We would like to thank Wing Suen, Yi Qian, Jonathan Kolstad, Hanming Fang, Jasmine Hao, Michael Wong,  and the seminar participants at Melbourne Institute: Applied Economic \& Social Research, and the applied economics workshop for helpful discussions to improve the paper. Xu acknowledges the general financial support by HKU Business School Shenzhen Center. Of course, all errors remain ours to bear.}
\author{Dan Jia}
\address{Bay Area International Business School, Beijing Normal University}
\email{jiadan@bnu.edu.cn}
\author{Pai Xu}
\address{HKU Business School, University of Hong Kong}
\email{paixu@hku.hk}

\begin{document}
	
\begin{abstract}
{\footnotesize }\bigskip\
The increasing prevalence of chronic diseases poses a significant challenge to global efforts to alleviate poverty, promote health equity, and control healthcare costs. This study adopts a structural approach to explore how patients manage chronic diseases by making trade-offs between inpatient care and ambulatory care outpatient services. Specifically, it investigates whether disadvantaged populations make distinct trade-offs compared to the general population and examines the impact of anti-poverty programs that reduce inpatient cost-sharing. 
Using health insurance claims data from a rural county in China, the study reveals that disadvantaged individuals tend to avoid ambulatory care unless it substantially lowers medical expenses. In contrast, the general population is more likely to prioritize ambulatory care, even at higher costs, to prevent disease progression. The findings also indicate that current anti-poverty insurance policies, which focus predominantly on hospitalization, inadvertently decrease ambulatory care usage by 23\%, resulting in increased healthcare costs and a 46.2\% decline in patient welfare. Counterfactual analysis suggests that reducing cost-sharing for ambulatory care would be a more cost-effective strategy for improving health outcomes and supporting disadvantaged populations than providing travel subsidies.
		
{\footnotesize \ }
		
 \bigskip
		
\end{abstract}
	
\date{\today. \;First version: March 2024.}
\maketitle

\section{Introduction}
Chronic diseases stand as leading causes of death and major contributors to healthcare costs worldwide. The escalating prevalence of chronic diseases not only increases healthcare expenditures but also poses a significant challenge to global efforts in reducing poverty and enhancing health equity. Disadvantaged populations bear a disproportionate burden, experiencing worse health outcomes, higher mortality rates, and greater financial hardship. \footnote{See, for example, \href{https://www.who.int/news-room/fact-sheets/detail/noncommunicable-diseases}{the Fact sheet of noncommunicable diseases} for details.} 

To mitigate these challenges, many nations have implemented health insurance programs designed to alleviate the financial burden on disadvantaged groups by expanding access and reducing cost-sharing, particularly for hospital-based care. For instance, in the United States, Medicare provides health insurance for individuals aged 65 or older and for younger individuals with disabilities, while Medicaid, expanded under the Affordable Care Act, supports low-income populations. Similarly, China’s resident insurance system extends coverage to rural populations and those outside the formal employment sector, offering substantially reduced cost-sharing for disadvantaged groups.

However, the design of these programs may lead to unintended consequences due to cross-price effects. Reducing cost-sharing for inpatient care can inadvertently discourage the use of other essential medical services. For example, \cite{ChandraAER2010} demonstrated that increasing cost-sharing for outpatient care led to fewer physician visits but higher rates of hospitalization. Consequently, anti-poverty programs focused primarily on lowering inpatient care costs may disincentivize ambulatory care, which is vital for early detection and timely treatment of chronic diseases. This shift can ultimately increase overall healthcare expenditures and worsen health outcomes for disadvantaged populations. Moreover, individuals from different socioeconomic backgrounds may respond differently to changes in healthcare prices. Disadvantaged individuals, in particular, may make distinct trade-offs between inpatient and ambulatory care, leaving them more vulnerable to the adverse effects of such policies.

In this paper, we address two key questions: First, do disadvantaged individuals make different trade-offs between inpatient and ambulatory care in managing chronic diseases compared to the general population? Second, how do anti-poverty programs that reduce inpatient cost-sharing affect healthcare utilization and costs of disadvantaged populations? Additionally, we explore alternative policy tools that may better support these populations while promoting health equity and cost-effectiveness.

We obtain and study a unique dataset of health insurance claims from a rural county in Western China to explore these issues. The data consists of medical records for outpatient and inpatient services for all patients enrolled in the resident insurance scheme between 2018 and 2020. This public insurance scheme covers more than 70\% of the national population and  99\% of the eligible population in the county. Each claim contains detailed medical information, such as the hospital, diagnosis, total expenditure, and out-of-pocket spending. Additionally, we can observe patient characteristics, including demographics, residence, and participation in any welfare assistance programs. 

Several additional features of China's healthcare system create a suitable context for analyzing patient demand in chronic disease management, and especially for contrasting the disadvantaged and the relatively well-off. Patients in China can directly access hospital care, including outpatient and inpatient services, without referrals. This eliminates the possibility that disadvantaged patients use ambulatory care solely due to limited access to specialists. Since ambulatory care is delivered through outpatient services in hospitals, there are minimal quality differences between ambulatory care and other types of hospital care. Finally, since physicians and hospitals are paid under a fee-for-service model this makes potential discrimination against disadvantaged patients, who often have higher risks, less likely.\footnote{It is worth noting that when physicians are paid under a capitation model, where payments are determined by the number of residents regardless of utilization, they may prefer to provide care to low-risk patients.}

We focus on patients with cardiovascular diseases (CVDs) because they serve as an ideal study group to examine the impact of inpatient cost-sharing on ambulatory care utilization. CVDs are responsible for the highest number of deaths and the largest economic burden among all chronic diseases. Furthermore, CVDs are considered ambulatory care-sensitive conditions (ACSCs), meaning that timely outpatient care effectively decreases the risk and severity of subsequent hospitalization. In this county, patients enrolled in a chronic disease care program use outpatient services to manage hypertension and diabetes as part of their CVD treatment. For this study, we categorize these services as ambulatory care.

In our data application, we observed that only disadvantaged patients are more likely to use ambulatory care as their disease severity increases, in sharp contrast with the non-disadvantaged. The most significant difference in ambulatory care use is among patients with mild conditions, with this gap narrowing as the severity of illness increases. Accordingly, we build a model to explain a patient's utilization choices. 

In our framework, patients weigh the trade-off between using ambulatory care early versus opting directly for inpatient care. We conceptualize individuals as having two primary motivations for accessing ambulatory care: reducing medical costs and preventing disease progression. Patients may prioritize disease prevention and financial savings differently, leading to distinct trade-offs between inpatient and ambulatory care. To capture these differences, we define a weighting parameter that represents the maximum additional medical expense a patient is willing to incur to maintain their current health condition. We anticipate that this willingness to bear extra costs will vary across different populations.


Using a binomial logit model, we estimated the parameters of this framework through maximum likelihood estimation. Our findings suggest that, on average, regular patients are willing to pay an additional RMB 725 (equivalent to USD 110) to preserve their current health by using ambulatory care. In contrast, disadvantaged patients only use ambulatory care if their medical costs can be reduced by at least RMB 535 by doing so. This indicates that they are less likely to seek preventive care unless the financial savings are substantial. The difference in weighting parameter helps explain why the gap in ambulatory care usage between disadvantaged and regular patients narrows as the severity of illness increases.

We then examine how anti-poverty programs impact disadvantaged patients by altering their inpatient cost-sharing. As part of China’s anti-poverty campaign, the local government in our sample county introduced a special insurance plan to make healthcare more affordable for disadvantaged populations. This plan provided increased financial support for hospitalizations, reducing their cost-sharing burden. In 2020, a further anti-poverty policy was implemented, lowering inpatient cost-sharing for disadvantaged patients by more than 40\%. We thus exploit the 2020 policy change to analyze its effects on the ambulatory care utilization of disadvantaged patients, employing a difference-in-difference approach. Regular patients, whose cost-sharing remained relatively unchanged, serve as the control group. Our findings indicate that reducing inpatient cost-sharing led to a 6 percentage point decrease in ambulatory care use among disadvantaged patients, with a particularly pronounced reduction for patients with more severe conditions.

Finally, we conduct several experiments to evaluate the effectiveness of various policy interventions for assisting disadvantaged populations. Our counterfactual analysis reveals that the current anti-poverty insurance plan makes disadvantaged patients 23.1\% less likely to use ambulatory care, which increases overall medical costs by RMB 290 (USD 44) and results in a welfare loss of 46.2\% per patient. Additionally, we find that policies reducing ambulatory care cost-sharing directly are more effective than those offering travel subsidies for ambulatory services. Specifically, lowering ambulatory care cost-sharing by 20 percentage points increases its utilization by 16.9\% and boosts patient welfare by 38.8\%. Implementing this policy would cost an average of RMB 32 (USD 5) per patient, while the estimated reduction in overall medical costs would exceed RMB 202 (USD 31).

	\cite{FinkelsteinQJE2016}
	
\textbf{The literature.} \quad This paper contributes to two strands of literature. The first is on healthcare demand. Recent studies have highlighted the importance of demand-side factors in explaining geographic variation in healthcare utilization. For example, \cite{FinkelsteinQJE2016} leveraged patient migration data to differentiate between supply-side and demand-side effects, discovering that half of the utilization differences in the US could be attributed to patient health status and demographic characteristics. Similarly, \cite{SalmJHE2020} reported an even greater influence of patient characteristics on ambulatory care utilization in Germany. 

As a result, anti-poverty programs aimed at increasing the number of healthcare providers or offering more flexible services to improve access for disadvantaged populations may be ineffective if patients are not motivated to use the services provided. \cite{BaileyAER2015} explored the establishment of Community Health Centers (CHCs) in the U.S. and found that while expanding primary care supply through CHCs reduced mortality rates in urban areas, it had less significant effects on non-white populations. This suggests that understanding patient demand is crucial for designing policies that effectively support disadvantaged individuals.

Patients typically consider a trade-off between benefits and costs in healthcare demand models. The benefits of healthcare can include improved ability to work and, consequently, increased wage income (\cite{GrossmanJPE1972}, and \cite{GilleskieETMA1998}). In contrast, cost concerns in shaping healthcare demand may arise from health insurance and time costs. (See, for example, \cite{ManningAER1987}, \cite{LuciforaHE2018}, \cite{FinkelsteinQJE2016}.) Within this framework, recent studies have increasingly delved into behavioral factors to understand patients' under-utilization of high-value preventive care. \cite{BaickerQJE2015} discussed how preferences for immediate costs over delayed benefits, inattention to non-salient symptoms, and biased beliefs can affect healthcare decision-making. 


Within this cost-benefit trade-off framework, our paper contributes by examining heterogeneity in how different populations evaluate treatment benefits. By analyzing variations in patients' weighting parameters when balancing inpatient and ambulatory care, we assess how anti-poverty programs that reduce inpatient cost-sharing impact the use of preventive care and healthcare costs for disadvantaged individuals. Our study provides important policy insights for encouraging the use of high-value care among disadvantaged populations, with broader implications for countries striving to reduce health disparities.

Secondly, this work contributes to the literature on the impacts of health insurance programs. Public health insurance initiatives aimed at expanding coverage or reducing cost-sharing have been implemented globally to make healthcare more affordable for disadvantaged populations. For instance, \cite{KolstadJPubE2012} and \cite{MillerJPubE2012} examined the 2006 Massachusetts health reform, finding that increased insurance coverage resulted in fewer non-urgent emergency visits and preventable hospital admissions, promoting more appropriate use of medical services.

Recent studies suggest that raising patient cost-sharing can reduce medical spending, but it may also lead to improper use of healthcare services. For example, \cite{ChandraAER2010} found that higher outpatient care cost-sharing increased hospitalizations, while \cite{BrotGoldbergQJE2017} showed that although higher cost-sharing lowered overall healthcare spending, it also reduced the use of high-value services such as preventive care and chronic disease management.

Our study extends the existing literature by examining the effects of cost-sharing when patients weigh different medical services in distinct ways. We provide empirical evidence that lowering cost-sharing for high-value services, such as ambulatory care for chronic diseases, can reduce overall healthcare costs, particularly for disadvantaged patients who tend to delay care and present with more severe conditions. These findings have important implications for anti-poverty insurance policies, highlighting that subsidizing high-value care is a more effective strategy for encouraging appropriate healthcare utilization and reducing costs.


The remainder of this paper is organized as follows: Section 2 provides the institutional background, while Section 3 discusses the data and presents the reduced-form evidence. Section 4 outlines the model, followed by the estimation approach and results in Section 5. Section 6 presents the counterfactual analysis, and the paper concludes in the final section.

\section{Background}
\subsection{Managing Chronic Diseases}
Chronic diseases are considered as the leading causes of death and the major drivers of healthcare costs across countries. The World Health Organization (WHO) recommends using primary healthcare for early detection and timely treatment as the critical approach to managing chronic diseases.\footnote{See, for example, \href{https://www.who.int/news-room/fact-sheets/detail/noncommunicable-diseases}{the Fact sheet of noncommunicable diseases} for details.} Academic studies have documented such intervention are cost-effective. For example, using primary care is shown to be associated with lower hospitalization rates and healthcare costs. (See, \cite{StarfieldNEJM2008, StarfieldMQ2005}.)
	
Many developed countries have developed strong systems of primary healthcare to manage chronic diseases in general population. Their general practitioners or family doctors  provide prevention-oriented care in communities. These countries also offer universal health insurance to all residents. Therefore, managing long-term chronic diseases is both accessible and affordable in these countries. 
	
In contrast, however, primary healthcare services in the US are much less comprehensive and underused when compared with other developed countries. (See, for example, \cite{StarfieldNEJM2008}.) With the absence of universal health insurance, the availability of health management to chronic diseases is greatly reduced. In the fight for affordable and comprehensive care to the under-served population, US government established community health centers (CHCs) with federal funds. These CHCs are located in the disadvantaged neighborhoods and offer almost-free medicine and examinations. \cite{BaileyAER2015} found that the establishment of Community Health Centers (CHCs) led to significant and sustained reductions in cardiovascular-related mortality rates among adults aged 50 and older. However, the impact was notably less pronounced in rural areas and among non-white populations. These findings suggest that focusing solely on the supply side may not provide a complete understanding of chronic disease management.

	
China, like many other developing countries, has a less established system for providing primary healthcare services. On one hand, China faces growing burden of aging society and chronic diseases. On the other hand, China's public hospitals are only major providers for most medical services and healthcare needs. Consequently, an enormous amount of health expenditure has been devoted to strengthen the local clinics in order to help the patients with chronic conditions managing their health.\footnote{The Chinese government emphasized the importance of primary healthcare in enhancing public health and controlling healthcare costs through the ``Healthy China 2030'' Initiative. In response, the State Council has developed a corresponding plan for the prevention and treatment of chronic diseases. See \href{http://www.gov.cn/zhengce/content/2017-02/14/content_5167886.htm}{China's Medium- and Long-term Plan for Prevention and Treatment of Chronic Diseases (2017-2025)} for details.} 
	
Specifically, township health centers in rural areas are tasked with screening, early detection, and health management of hypertension and diabetes. Long-term care for chronic diseases is also provided through outpatient ambulatory services at public hospitals, with these services typically covered by government health insurance. However, as seen in the U.S. healthcare market, it is uncertain whether expanding ambulatory care for chronic disease management will effectively improve population welfare. Ultimately, the decision to utilize such services to prevent or reduce hospitalizations rests with the patients themselves. Therefore, in this study, we will investigate how patients trade off between ambulatory and inpatient care when managing chronic diseases.

\subsection{Insurance Schemes} \label{data}
There are two public insurance schemes in China: employment insurance and resident insurance. A resident under formal employment is mandated to have  employment insurance. This is registered and enforced with salary payment. All other residents, rural or urban, are eligible for resident insurance. In 2018, public insurance covered more than 95\% of the national population, with resident insurance accounting for around 76.5\% of those covered.\footnote{Source: statistical bulletin of National Healthcare Security Administration.} 
	
Our dataset covers the insurance claim records for all patients who enrolled in the resident insurance scheme of the county from 2018 to 2020. Despite the voluntary nature, the resident insurance covers more than 99\% of the eligible population in the county. Our data can thus be regarded as a sample from a universal health insurance scheme.\footnote{Source: government work report of the county.} 
We will introduce the local resident insurance schemes in this subsection. All the local residents are eligible to enroll in a regular insurance plan with an annual premium RMB 250 (equivalent to USD 35). We list the cost-sharing rates under this regular plan in the column (1) of Table \ref{tab:ins/plan}.

\begin{table}[htbp]\centering
	\footnotesize
\def\sym#1{\ifmmode^{#1}\else\(^{#1}\)\fi}
\caption{Insurance Plans \label{tab:ins/plan}}
\begin{threeparttable}
\begin{tabular}{l*{2}{c}}
\hline\hline
  &\multicolumn{1}{c}{(1)}         &\multicolumn{1}{c}{(2)}            \\
                &\multicolumn{1}{c}{Regular}&\multicolumn{1}{c}{Poor}\\
\hline
Annual Premium      &    \$35 & -  \\
\hline
\emph{Panel A. Ambulatory Care} & & \\
Deductible  &   -  &  -    \\
Coinsurance & 30\% & 30\%  \\
Assistance & - & 50\%  \\
Ave Cost-sharing & 35.7\% & 33.7\%  \\
\hline
\emph{Panel B. Inpatient Care} & & \\
Deductible  &   \$7.5/\$45/\$135 (visit) &  -    \\
Coinsurance & 20\%/25\%/50\% & 15\%/20\%/45\% \\
Assistance & - & 80\%  \\
Ave Cost-sharing & 49.7\% & 29.3\%  \\
\hline\hline
\end{tabular}
\end{threeparttable}
\end{table}

When a patient manages the chronic conditions with ambulatory care, the plan sets the coinsurance rate uniformly at 30\% with no deductible. However, the coinsurance rates vary depending on the treatment facility types, if the inpatient care is needed. Specifically, the coinsurance rates are 20\%, 25\%, and 50\% for being hospitalized at township health center, secondary hospital and tertiary hospital, respectively. Moreover, the deductibles for these inpatient services also differ, ie, RMB 50 (equivalent to \$7.5), RMB 300 (equivalent to \$45), and RMB 900 (equivalent to \$135), respectively. 
In summary, the residents under the regular plan pay 35.7\% of their total medical bills for ambulatory care, and 49.7\% for their inpatient care. 

As a matter of fact, the county in this study is among the targets of China's anti-poverty campaign. Therefore, the local government has implemented various measures in assisting the disadvantaged population in the community. For example, the health insurance scheme introduced a special plan aiming to offer affordable healthcare to the poor household whose annual income is less than RMB 3000 (equivalent to US\$ 452). We list the parameters for this special plan in the column (2) of Table \ref{tab:ins/plan}.

The residents in these poor households need to pay no premium to participate in this special plan. There is also no deductibles for any medical visits. The coinsurance rates are 5\% less than the regular plan for treatment at any facility. Moreover, they receive additional financial aids to all the out-of-pocket spending for medical services, ie 50\%  for ambulatory care, and 80\% for inpatient care. In the end, the residents with financial stress pay 33.7\% on the total cost of their ambulatory care, but only 29.3\% of the medical bills when hospitalized.

\section{Data and Reduced-form Evidence}
	
Our dataset consists of health insurance claims from a rural county in western China, covering all patients enrolled in the county's resident insurance scheme from 2018 to 2020. Each claim provides detailed medical information, including the hospital, diagnosis, total expenditure, and out-of-pocket spending. Additionally, we have access to patient characteristics such as demographics, residence, and participation in welfare assistance programs.

\begin{figure}
		\includegraphics[width=0.45\textwidth]{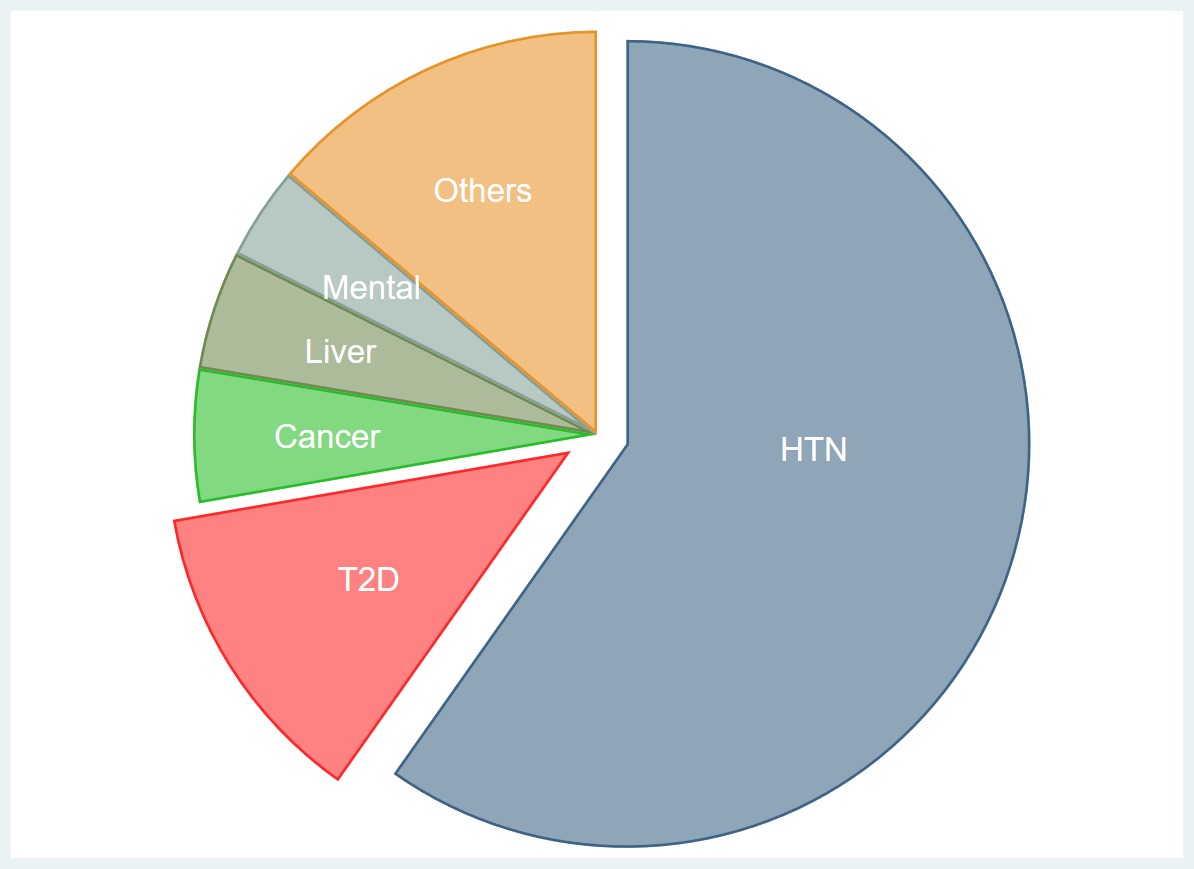}
		\caption{Ambulatory Care Utilization by Diagnoses}
		\label{Fig:chrout}
	\end{figure}

Patients who need managing chronic conditions, are supervised under a designated program in this county - Chronic Disease Care Program (CDCP). They became participants to the program once they are diagnosed with a disease on the predetermined list. Their long-term care are provided through the outpatient ambulatory services by three township health centers and the general hospital. Other facilities including the TCM are not qualified to offer such services in the county. We show the ratio of patient visits under this program by each diagnosis in Figure \ref{Fig:chrout}.  Among these, hypertension (HTN) and diabetes (T2D) are the most prevalent, accounting for 59.8\% and 12.5\% of the claimed cases, respectively. 
	
\subsection{Sample}	
In this paper, we investigate patients with cardiovascular diseases (CVDs) to study their ambulatory care utilization. The most common CVDs include coronary heart disease and cerebrovascular diseases such as stroke. We do so for two reasons. First, timely ambulatory care on CVDs can effectively reduce the risk and severity of subsequent hospitalization. Second, these patient cases not only occur with high frequency in the general population, but also account for most deaths universally. In comparison to other chronic diseases, CVDs have become the largest economic burden to healthcare systems in many nations around the globe. 
	
Managing risk factors of CVDs reduces the probability of their severe outcomes.  According to medical studies, hypertension and diabetes are the most important conditions to take care in managing CVDs. Therefore, we consider patients who have been treating these two diagnoses under the CDCP program as having taken the primary care to CVDs. 
	
For the purpose of our study, we focus on the patients who have been diagnosed and hospitalized by CVD conditions. It is unambiguous that these patients have been notified their conditions. They are also aware the option of ambulatory care that can definitely help improve their health conditions. This criterion selected 685 patients and 2180 hospitalization records from the data. The most common CVDs in hospitalization include cerebral infarction, intracerebral hemorrhage and coronary heart disease. 
	
Hypertension, diabetes or CVDs rarely occur at young ages in this county.  When we incidentally observe young patients in the data, we are more inclined to think some unobserved characteristics in their health played a role. Thus, suggesting primary care usage may have a different implication for patients whose age group typically anticipates CVD occurrences than for them.  We further exclude patients under age 40 from our sample.  Therefore, our working sample includes 630 patients and 2023 hospitalization records.

\subsection{Descriptive Statistics} \label{empirical} 
In this subsection, we make a comparison between the patients who have used ambulatory care and who have not in the working sample. The results are reported in Table \ref{Tab:data/ss}. First, Panel A compares the two groups of patients on their observed characteristics. Columns (2) and (3) list the group mean on each dimension, respectively. Columns (4) presents their difference. 
	
\begin{table}[htbp]\centering
	\footnotesize
\def\sym#1{\ifmmode^{#1}\else\(^{#1}\)\fi}
\caption{Patients With and Without Ambulatory Care\label{Tab:data/ss}}
\begin{threeparttable}
\begin{tabular}{l*{4}{c}}
\hline\hline
                    &\multicolumn{1}{c}{All}&\multicolumn{1}{c}{With Care}&\multicolumn{1}{c}{W/O Care}&\multicolumn{1}{c}{Difference}\\
                    &\multicolumn{1}{c}{(1)}         &\multicolumn{1}{c}{(2)}         &\multicolumn{1}{c}{(3)} &\multicolumn{1}{c}{(4)}        \\
\hline
\emph{Panel A. Patient Characteristics}      &     &      &   & \\
Age                 &       69.29&       71.20&       68.47&       -2.73\sym{**}\\
Male                &       0.475&       0.392&       0.510&        0.12\sym{**}\\
Minority            &       0.095&       0.058&       0.111&        0.05\sym{*}\\
Urban               &       0.342&       0.418&       0.309&       -0.11\sym{**}\\
Travel distance     &      11.089&       9.553&      11.759&        2.21\\

Disadvantaged       &       0.748&       0.629&       0.798&        0.17\sym{***}\\
\quad Low-income    &       0.489&       0.444&       0.508&        0.06\\
\quad Distant       &       0.525&       0.423&       0.569&        0.15\sym{***}\\

&&&&\\
\emph{Panel B. Disease Severity}      &     &      &  &  \\
Discrete      &             &            &          &   \\
\quad Mild    &        0.394&       0.344&       0.415&        0.071\\
\quad Moderate&        0.465&       0.508&       0.447&       -0.061\\
\quad Severe  &        0.141&       0.148&       0.138&       -0.010\\
Preference-discounted     &        0.393&       0.384&       0.397&        0.013\\
&&&&\\
\emph{Panel C. Hospitalization Cost}          &     &      & &   \\
Type 1: Clinic  &     4,654&     3,498&     5,016&    1518\sym{*}\\
Type 2: Hospital   &     8,126&     7,364&     8,469&        1104\\
Type 3: Non-local   &    27,480&    26,445&    28,060&        1615\\
\hline
Observations        &         834&         261&         573& \\
No. of patients        &         630&         189&         441& \\
\hline\hline
\end{tabular}
\begin{tablenotes}[flushleft]\footnotesize
\note The observation in Panel C is patient-year CVD hospitalizations at each facility type. They are in RMB. 
\end{tablenotes}
\end{threeparttable}
\end{table}

	
%
Our data set comes from a county covers remote villages and disadvantaged population under various welfare programs. Therefore, we check whether a patient belongs to ``low-income" as if (s)he participated in special welfare programs: receiving minimum living allowance from government or his/her household identified as poor for assistance. We also check whether a patient's residential address is far away from the county center.  ``Distant" is one if the patient has to travel more than 12 km to the county center, and zero otherwise.\footnote{This 12-km threshold distinguishes townships located in the same valley with the county center and townships cut off by mountains, as shown in appendix \ref{map}. As a matter of fact, patients averagely travel more than 20 km to the county center and the farthest need to travel more than 90 km. } In this paper, individuals with low income or residing in distant villages are categorized as disadvantaged. 

Furthermore, we examine other characteristics most relevant to this application. If a patient resides in a village associated with a Tibetan tribe, we classify them as ``Minority''. Members of this group share a common language, cultural practices, and social norms. Additionally, we verify if their residences are situated within an officially designated urban area, which is indicated as ``Urban''. To assess the accessibility of ambulatory care for each patient, we utilize ``travel distance'' as a metric. This measure represents the number of kilometers a patient must travel to reach the nearest facility. We learn from Panel A that most of these factors matter to ambulatory care usage. Male Patients, minority patients, and those with financial stress and travel constraints are associated with less demand in ambulatory care. Instead, elder and urban residents exhibit a higher likelihood of opting for ambulatory care services.
	
Second, we consider disease severity in our analysis. However, a key challenge arises from the fact that patients' disease severity is not directly observable. To address this, we develop a measure of severity based on patients' annual hospitalization costs for conditions unrelated to CVDs. Our approach assumes that disease severity is positively correlated with medical costs and remains consistent across diagnoses.
	
We divide the patients in our data into three different levels of severity. In particular, the patient is considered as Mild if she has not been hospitalized for any other diagnosis. If her yearly average hospitalization cost for other diagnoses is under fifteen thousand RMB (equivalent to USD 2262), she has Moderate condition; otherwise, her condition is deemed Severe.\footnote{One concern is that patients who do not use inpatient care may imply preference instead of mild. Therefore, we exclude those patients in the robustness check. Our results remain robust.} Panel B of Table \ref{Tab:data/ss} presents the distribution of disease severity between patients who utilized ambulatory care and those who did not. The data indicate that a higher proportion of patients with mild conditions chose to forgo ambulatory care, although the difference is not statistically significant.
	
A potential critique of using hospitalization cost as a measure of severity is that its variation encompasses both health conditions and patients' preferences for care. In other words, patients who prefer higher quality or more intensive care may also choose to spend more on hospitalization, even under less severe conditions. To address this concern, we construct an alternative severity measure that adjusts for systematic effects related to patients' observed preferences for care, including demographics and facility choices.
	
The new preference-discounted measure is continuous, ranging from 0 to 1. We assume that any unobserved preferences remain constant across diagnoses. As a result, if a patient spends more on other diseases, it is likely that they would also spend more among the population of CVD patients, all other factors being constant. Our revised severity measure incorporates information from various types of diagnoses, weighting them based on their frequency. Further details on how we construct these severity measures can be found in the appendix.
	
Since the Mild patients have no hospitalization spending on other diagnoses, we thus exclude those patients for this preference-discounted measure. We observe no significant difference in disease severity between the groups with and without ambulatory care. This observation is consistent with using the discrete measure on severity. This seems to suggest that preference differences are not primary concern when measuring disease severity by hospitalization costs.

We next examine average hospitalization costs for CVDs at each facility type between two patient groups in Panel C of Table \ref{Tab:data/ss}. There are 18 township health centers throughout the rural areas in the county. They function as local clinics and provide basic services in the community for common diseases such as cold. In addition, two secondary hospitals are available in the county center, with one serving only traditional Chinese medicine (TCM). Even though graded as secondary, the TCM hospital should be effectively regarded as a local clinic as it offers only basic medical services. 

The best medical care available in the county is offered by its General Hospital, as there is no tertiary hospital in this region. However, medical conditions with complication or challenges can only be treated in tertiary hospitals. Local residents have to travel by more than 5 hours of driving to the nearest tertiary hospital for more advanced medical services.

We classify treatment facilities in our sample in accordance with their actual treatment capacity. ``Type 1 Clinic" includes township health centers (THC) and the traditional Chinese medicine hospital (TCM). The local general hospital is considered as ``Type 2 Hospital", and the non-local tertiary hospitals are labeled as ``Type 3 Non-local". The patients who have used ambulatory care spent less on hospitalization regardless the facility type visited. It is rather interesting to observe this, because these patients may have relatively worse disease severity overall as we learnt from Panel B. It suggests possible benefits ambulatory care may have offered. Moreover, hospitalization costs increase drastically over facility types.

\subsection{Different Utilization by the Disadvantaged}
In this subsection, we employ probit regressions to analyze patient choices regarding ambulatory care. Our key interest lies in discerning any distinct patterns of ambulatory care utilization among disadvantaged patients. To this end, we investigate the interaction effects between disease severity and the disadvantaged status of patients, while controlling for demographic variables and hospital type choice. 

\begin{table}[htbp]\centering
	\footnotesize
\def\sym#1{\ifmmode^{#1}\else\(^{#1}\)\fi}
\caption{Marginal Effects on Using Ambulatory Care \label{tab:use/pred}}
\begin{threeparttable}
\begin{tabular}{l*{3}{c}}
\hline\hline
                &\multicolumn{3}{c}{Prob. of Ambulatory Care}            \\\cmidrule(lr){2-4}
                &\multicolumn{1}{c}{(1)}         &\multicolumn{1}{c}{(2)}         &\multicolumn{1}{c}{(3)}         \\
\hline
Disadvantaged   &   -0.265\sym{***}&   -0.255\sym{***}&   -0.251\sym{***}\\
                &  (0.064)         &  (0.069)         &  (0.069)         \\
Disadvantaged $\times$ Moderate&    0.113         &    0.142         &    0.150\sym{*}  \\
                &  (0.087)         &  (0.089)         &  (0.089)         \\
Disadvantaged $\times$ Severe&    0.243\sym{*}  &    0.259\sym{*}  &    0.271\sym{*}  \\
                &  (0.134)         &  (0.138)         &  (0.139)         \\
Moderate        &   -0.003         &   -0.032         &   -0.043         \\
&  (0.073)         &  (0.074)         &  (0.074)         \\
Severe          &   -0.110         &   -0.121         &   -0.136         \\
&  (0.118)         &  (0.120)         &  (0.121)         \\                
Travel distance&                  &    0.000         &    0.000         \\
                &                  &  (0.001)         &  (0.001)         \\
Age             &                  &    0.005\sym{***}&    0.006\sym{***}\\
                &                  &  (0.002)         &  (0.002)         \\
Male            &                  &   -0.098\sym{***}&   -0.096\sym{**} \\
                &                  &  (0.038)         &  (0.038)         \\
Urban &                  &    0.063         &    0.059         \\
                &                  &  (0.044)         &  (0.044)         \\
Minority     &                  &   -0.123\sym{*}  &   -0.119\sym{*}  \\
                &                  &  (0.072)         &  (0.072)         \\
Type 2: Hospital&                  &                  &    0.069         \\
                &                  &                  &  (0.052)         \\
Type 3: Nonlocal&                  &                  &    0.123\sym{**} \\
                &                  &                  &  (0.059)         \\
\hline
Observations    &      630         &      622         &      622         \\
Adjusted \(R^{2}\)&                  &                  &                  \\
\hline\hline
\end{tabular}
\begin{tablenotes}[flushleft]
\note Standard errors are in parentheses. We use patient level data and probit models for this analysis. Travel distance measures the travel distance to the nearest ambulatory care provider and is in the unit of every twelve kilometers. \sym{*} \(p<.10\), \sym{**} \(p<.05\), \sym{***} \(p<.01\).
\end{tablenotes}
\end{threeparttable}
\end{table}

We first find that disadvantaged patients are approximately 28\% less likely to use ambulatory care compared to regular patients. However, this difference in utilization decreases as the severity of the disease worsens. Additionally, older patients, female patients, non-minority patients, and those who choose higher-quality hospitals are more likely to use ambulatory care. Lastly, travel distance, which measures accessibility to ambulatory care, does not significantly influence its utilization, suggesting that other demand-side factors may play a more important role.
	
\begin{figure}
	\includegraphics[width=0.55\textwidth]{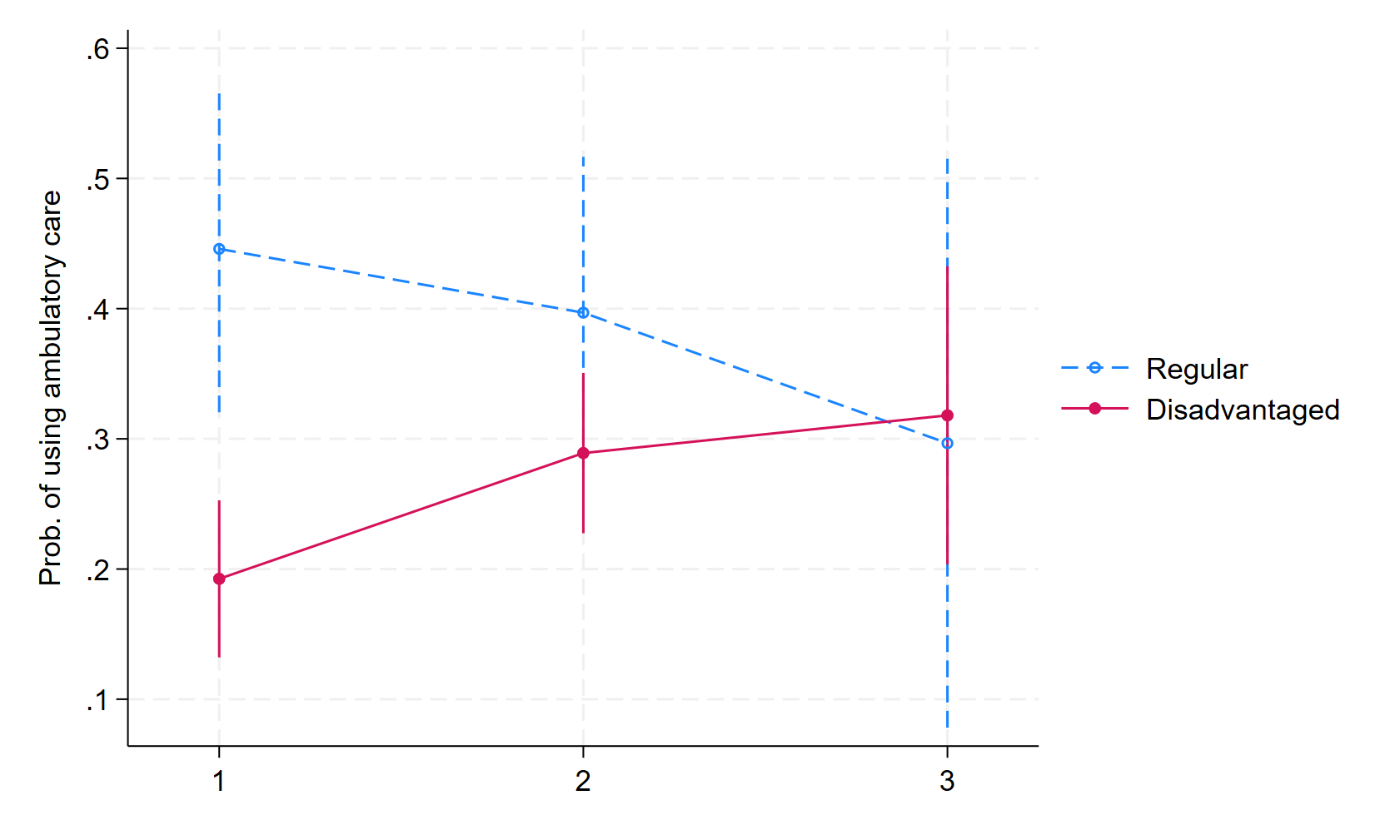}
	\caption{Predicted Probability of Using Ambulatory Care}
	\label{Fig:use/pred1}
\end{figure}
	
We then plot the predicted probability of using ambulatory care between disadvantaged patients (red solid line) and others (blue dashed line) at each level of disease severity in Figure \ref{Fig:use/pred1}. It shows a most significant difference in likelihood of using ambulatory care for patients with mild severity. However, this difference diminishes notably as the severity of patients' conditions worsens. Consequently, disadvantaged patients do not exhibit lower utilization of ambulatory care services compared to others when their conditions become severe.

Sicker patients typically have a greater need for medical services, which drives higher demand for healthcare. This health-driven demand explains why those in more severe conditions are more likely to utilize ambulatory care, as the benefits in reducing hospitalization costs are more pronounced. At the same time, ambulatory care can also benefit patients with milder conditions by preventing their diseases from worsening, offering an additional incentive to seek such care.

The observed difference in utilization between disadvantaged and regular patients suggests they may prioritize treating current conditions and disease prevention differently. Specifically, disadvantaged patients might be more focused on financial savings and therefore may only use ambulatory care when their conditions are severe enough that it can significantly reduce inpatient costs. In contrast, regular patients may prioritize disease prevention and begin using ambulatory care even when their conditions are mild. In the next section, we will develop an economic model to explore this trade-off further.

\subsection{Effects of Anti-poverty Insurance Policy}
In this subsection, we analyze the impact of an anti-poverty program on the use of ambulatory care among disadvantaged patients by leveraging an exogenous policy shock that changed their inpatient cost-sharing.

As a continuing effort to fight poverty in rural China, the national campaign further empowered its assistance to disadvantaged populations in 2020. Being part of this effort, more generous support for hospitalization is provided in order to increase affordability of the disadvantaged. Particularly in this county, the residents with low income levels or living distant from the county center all benefited from the expanded welfare programs. The new healthcare plan effectively reduced the cost-sharing for disadvantaged patients, when they are hospitalized for treatment. 

We take advantage of this exogenous policy change and examine how healthcare utilization may change accordingly. In particular, we wonder whether more generous support to inpatient care has impacted ambulatory care utilization at all. To this end, we first show the changes in the average cost-sharing for inpatient care before and after the insurance policy change in Table \ref{Tab:ss/ciptyl}. It shows that, regardless of facility type, the cost-sharing of the disadvantaged patients is reduced by more than 40\%. Instead, the regular residents only can benefit with less cost-sharing when they visit non-local hospitals. 

\begin{table}[htbp]\centering
	\footnotesize
\def\sym#1{\ifmmode^{#1}\else\(^{#1}\)\fi}
\caption{Average Cost-sharing for Inpatient Care\label{Tab:ss/ciptyl}}
\begin{threeparttable}
\begin{tabular}{l*{6}{c}}
\hline\hline
                    &\multicolumn{3}{c}{Regular Residents}&\multicolumn{3}{c}{Disadvantaged}\\\cmidrule(lr){2-4}\cmidrule(lr){5-7}
                    &\multicolumn{1}{c}{Before}&\multicolumn{1}{c}{After}&\multicolumn{1}{c}{Difference}&\multicolumn{1}{c}{Before}&\multicolumn{1}{c}{After}&\multicolumn{1}{c}{Difference}\\
\hline
Type 1: Clinic             &       0.304&       0.319&      -0.015&       0.235&       0.124&       0.111\sym{***}\\
Type 2: Hospital       &       0.288&       0.293&     -0.005 &   0.239&       0.141&       0.098\sym{***}\\
Type 3: Non-local             &       0.513&       0.390&       0.123\sym{***}&       0.494&       0.269&       0.224\sym{***}\\
\hline\hline
\end{tabular}
\begin{tablenotes}[flushleft]
\note ``Before" indicates the hospitalization during 2018-2019; ``After" indicates hospitalization in 2020.
\end{tablenotes}
\end{threeparttable}
\end{table}

We next consider the effect on ambulatory care utilization by the policy change. Since the regular residents did not experience much of benefits from the policy change, we use them as the control group. Our focus is thus on the disadvantaged patients and their decisions on using ambulatory care. We employ a difference-in-difference approach and use the disadvantaged as the treatment group. We use probit model to examine how reducing cost-sharing in inpatient care may affect utilizing the ambulatory care. The marginal effects of the policy are reported in Table \ref{tab:use/insptyl}.

\begin{table}[htbp]\centering
	\footnotesize
\def\sym#1{\ifmmode^{#1}\else\(^{#1}\)\fi}
\caption{Marginal Effects on Using Ambulatory Care \label{tab:use/insptyl}}
\begin{threeparttable}
\begin{tabular}{l*{3}{c}}
\hline\hline
                &\multicolumn{3}{c}{Probability of Ambulatory Care}                    \\\cmidrule(lr){2-4}
                &\multicolumn{1}{c}{(1)}         &\multicolumn{1}{c}{(2)}         &\multicolumn{1}{c}{(3)}         \\
\hline
Disadvantaged $\times$ Post&   -0.060\sym{*}  &   -0.061\sym{**} &   -0.040         \\
                &  (0.031)         &  (0.031)         &  (0.033)         \\
Year Fixed Effects & Yes & Yes & Yes\\                
Demographics    & & Yes & Yes \\ 
Disease Severity &  & Yes & Yes\\            
Mild/Moderate Only & & & Yes \\
\hline
Observations    &     1890         &     1890         &     1623         \\
\hline\hline
\end{tabular}
\begin{tablenotes}[flushleft]
\note We use patient-year data and probit models for this analysis. 
\end{tablenotes}
\end{threeparttable}
\end{table}

In general, less cost-sharing of inpatient care makes the disadvantaged patients less likely to use ambulatory care by around 6 percentage points. Such a magnitude is about 24.3\% of the average level. The result remains robust even if we take into account the patients' demographics and disease severity. (See column (2).) It seems to suggest that lower cost-sharing of inpatient care lead to less ambulatory care. One possible reason for this argument is that the ambulatory care becomes more expensive in comparison to inpatient care. The literature has documented similar cross-price effects. For example, \cite{ChandraAER2010} found that increasing the cost-sharing of outpatient care raised hospitalization utilization.

It is also interesting to observe that the effect of cost-sharing becomes statistically insignificant if we restrict our sample by excluding Severe patients. We have a rather small sub-sample on disadvantaged patients with Severe condition. It limits our ability of explore more thoroughly within this sub-population. But a quick look at summary statistic reveals an astonished fact: there was 11\% of the disadvantaged patients with Severe condition choose to use ambulatory care services before the policy came into effect, while the proportion dropped to zero after the policy. All in all, the effects of health insurance can be heterogeneous across the population, for example, by their disease severity.

In summary, health insurance plays a significant role in shaping healthcare demand. Additionally, when examining ambulatory care utilization, it is important to consider the potential price effects driven by health insurance plans. In the following section, we will develop an economic model to explore how the relative costs of inpatient and ambulatory care influence demand for ambulatory services.

\section{ The Model}
In this section, we present a model that explains a patient's decision to utilize ambulatory care. The core of this decision lies in balancing the relative benefits and costs of seeking ambulatory care early versus opting directly for inpatient care. We will examine how this choice is influenced by the patient's trade-off between disease prevention and financial savings, as well as the role of insurance cost-sharing in shaping their decision.\footnote{We also discuss several alternative factors that may impact ambulatory care usage, including time preference, symptom salience, and biased beliefs in appendix \ref{alternative}.}

\subsection{Patient's Utility}
	
Patient $i$'s severity of disease is represented by $\theta_i$ continuously distributed on $(0,1)$. The larger $\theta$ indicates more severe conditions. Our objective is to examine the variations in ambulatory care decision-making among patients with differing degrees of disease severity. Consequently, we consider disease severity as a patient attribute that remains constant within our framework. Therefore, the utility from using ambulatory care relative to opting directly for inpatient care is
$$
U_i = \underbrace{[P_{i,0}^{hc}(\theta_i) - P_{i,1}^{hc}(\theta_i)] + \gamma_i \cdot (1-\theta_i)}_{\text{benefits of ambulatory care}} - \underbrace{[P_i^{pc}(\theta_i) + T_i]}_{\text{costs of ambulatory care}},
$$
where $P$ is the cost incurred for medical care which depends on disease severity $\theta$, the subscript 0 represents the situation without using ambulatory care, the superscript $hc$ denotes inpatient care, and the superscript $pc$ represents ambulatory care. Thus, $P_{i,0}^{hc}$ is the expected medical cost for hospitalization when patient $i$ opts out of ambulatory care. Instead, if patient $i$ has incurred cost for ambulatory care, ie $P_i^{pc}$, and used ambulatory care services, her expected hospitalization cost is $P_{i,1}^{hc}$. The term $[P_{i,0}^{hc} (\theta_i)- P_{i,1}^{hc}(\theta_i)]$, thus measures the savings in inpatient care as the patient $i$'s first benefit of using primary care.
	
We express a patient's concerns about preventing disease deterioration by the distance from her current severity to the worst scenario, $(1-\theta_i)$. In other words, using ambulatory care to prevent worse severity values $1-\theta_i$ for patient $i$. This approach eliminates the need to estimate the probability of transitioning between severity levels when assessing the value of disease prevention. We also conducted experiments using a two-period model, where we defined the probability of severity transitions to quantify the value of preventing deterioration. The results from this alternative approach align with our current model, indicating that the value of disease prevention decreases as the patient's condition worsens.
	
We then introduce a weighting parameter, $\gamma_{i}$, to represent how much a patient prioritizes disease prevention relative to the monetary cost of treatment. This parameter is expected to vary based on the patient's socioeconomic status. A patient's valuation of disease prevention may stem from various factors, including their awareness, knowledge, and local customs. These factors contribute to heterogeneity in healthcare decisions, particularly among disadvantaged populations. We will examine whether disadvantaged patients have different weighting parameters compared to regular patients, and how factors such as living in rural areas or minority villages influence these differences in the next section.
	
$T_i$ is the travel cost for using ambulatory care. We use this to represent any inconvenience and opportunity cost involved for a patient to undergo the ambulatory care service. Using ambulatory care will incur extra medical and travel costs $[P_i^{pc} + T_i]$, but reduce the hospitalization cost from $P_{i,0}^{hc}$ to $P_{i,1}^{hc}$. The patient chooses to use ambulatory care only when it brings positive utility. That is, the benefits of ambulatory care in reducing hospitalization cost and preventing deterioration outweigh its costs, ie, $U_i>0$.
	
A non-zero $\gamma$ changes the simple monetary benefit and cost balancing in patients' decision-making.  When $\gamma_{i} > 0$, the patient is concerned about avoiding worsening health conditions and is more likely to use preventive treatment, even if it comes with higher medical costs. In contrast, $\gamma_{i} < 0$ indicates that the patient experiences disutility from using preventive care to maintain their current health, leading them to avoid ambulatory care, even when the financial benefits outweigh the costs. 

\subsubsection{Costs of Medical Care}\label{cost}
We next introduce the dependence of patient's medical care cost on her disease severity. If the patient $i$ does not use ambulatory care, her medical cost of hospitalization is assumed to follow 
$$P_{i,0}^{hc} = E(P_{i,0}^{hc}|P_{i,0}^{hc}>0) = \theta_i^{\beta} \bar{K}^s,$$
where $\bar{K}^s$ denotes the maximum medical cost for treating in hospital $s$. We focus on the patients who have already been hospitalized for chronic conditions in this model. Therefore, the probability of being hospitalized is 1. $\beta >0$ measures the effects of disease severity on hospitalization cost. Again, a more severe patient spends more on treatment in hospitals. 
	
Prior to hospitalization, the patient can choose to use ambulatory care for preventive treatment, whose cost is determined by 
\[
P_i^{pc} = \theta_i^{\alpha} \bar{P},
\]
where $\bar{P}$ is the benchmark, ie, the highest possible treatment cost for ambulatory care. $\alpha>0$ shows how disease severity matters to the cost of using ambulatory care. 
	
We assume the ambulatory care is effective by reducing patient's disease severity from $\theta$ to $\lambda\theta$, where $0< \lambda <1$. Therefore, the medical costs of hospitalization for those choosing to use ambulatory care is $$P_{i,1}^{hc} = (\lambda\theta_i)^{\beta} \bar{K}^s,$$
	
After substituting $P_{i,0}^{hc}, P_{i,1}^{hc},$ and $P_i^{pc}$ and some simple calculation, the decision rule for using ambulatory care is specified as below:
\begin{equation} \label{decision}
	U_i = (1-\lambda^{\beta}) \theta_i^{\beta} \bar{K}^s + \gamma_i (1 - \theta_i) - (\theta_i^{\alpha}\bar{P} + T_i)  >0.
\end{equation}

\subsection{Heterogeneity in Weighting}
\begin{figure}
	\includegraphics[width=0.48\textwidth]{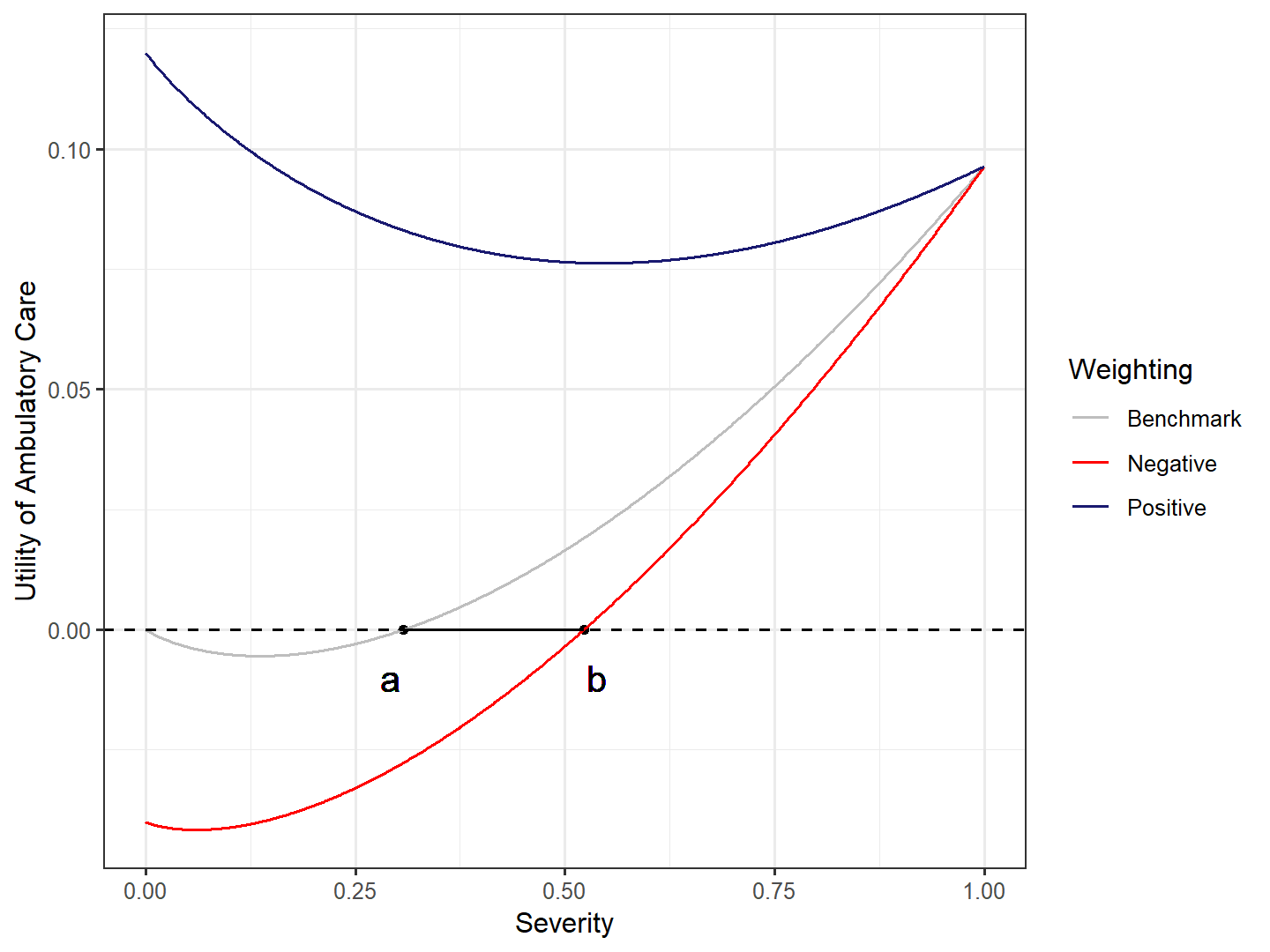}
	\caption{Utility of Using Ambulatory Care, effects of Weighting $\gamma$}
	\label{Fig:model/utha}
\end{figure}

We plot the utility of using ambulatory care over disease severity for different weighting parameter in Figure \ref{Fig:model/utha}. The gray curve represents the baseline case with $\gamma=0$, which determined by a simple monetary cost-benefit trade-off. The blue curve is for the positive weighting case where $\gamma=0.12$ and the red curve is for the negative case where $\gamma=-0.04$. Here we standardize the utility by dividing the maximal hospitalization cost, $\bar{K}^s$, to interpret the utility as the share of the maximal cost. The other parameters are specified as $\bar{P}/\bar{K}^s=0.12, \alpha=1, \beta=1.5, \lambda=0.85, T_i=0$. 
	
The patients with negative parameter would not go for ambulatory care until their condition turns more severe. On the contrary, the patients with positive parameter are most likely to put ambulatory care in demand when the disease is not severe.  Consequently, when patients prioritize financial savings to disease prevention, it is mainly the less severe ones discouraged away from using ambulatory care. This different patterns caused by weighting aligns with our observed disparity in ambulatory care usage between disadvantaged patients and others in our data (Figure \ref{Fig:use/pred1}). 

This different pattern towards disease severity comes from how the patient trades off health concerns $(1-\theta_i)$ and monetary benefits $(P_{i,0}^{hc} - P_{i,1}^{hc})$ in our model. When the weighting parameter is positive enough so that the health concern dominates, less severe ones will perceive larger utility from using ambulatory care. On the contrary, when the weighting parameter is negative enough so that the monetary benefit dominates, it is the more severe ones who perceive larger utility.

Moreover, for patients whose utility lies above the dashed black line, the use of ambulatory care results in positive financial savings. In other words, for any patient with a severity level worse than $a$, utilizing ambulatory care allows them to enjoy lower total costs for healthcare. However, patients with a negative weighting parameter will only opt for ambulatory care when their severity $\theta_i$ exceeds $b$. As a result, the negative weighting increases medical costs for patients with disease severity in the range of $[a,b]$.

\subsection{Effects of Health Insurance}
We then incorporate health insurance into our model. The key parameter for insurance plan is patient's cost-sharing, which mostly depends on the coinsurance rate in our application. But a critical feature is the cost-sharing for ambulatory care and inpatient care may be different. We denote the cost-sharing for ambulatory care and inpatient care by $\phi_{pc}$ and $\phi_{hc}$, respectively. The cost-sharing parameter comes along with the insurance plan, thus can be viewed as exogenous in our model. 
	
With health insurance, the patient's decision rule for using ambulatory care is modified as follows,
\begin{equation}\label{eq:pcdecisionwcpay}
	U_i = (P_{i,0}^{hc} - P_{i,1}^{hc}) + \gamma_i \frac{(1-\theta_i)}{\phi_{hc}} - (\frac{\phi_{pc} }{\phi_{hc}} P_i^{pc} + \frac{T_i}{\phi_{hc}})>0.
\end{equation}
In this case, the ratio of cost-sharing $\phi_{pc}/\phi_{hc}$ matters to the ambulatory care decision by changing the cost of ambulatory care. In order to examine the marginal effect of relative prices in healthcare, we consider a fixed cost-sharing for hospitalization, ie $\phi_{hc}$ is constant. Then, a marginal increase of cost-sharing for ambulatory care $\phi_{pc}$ becomes  $$-\theta_i^{\alpha} \bar{P}.$$ 
	
Firstly, this marginal effect is always negative as $\theta_i^{\alpha}>0$ and $\bar{P}>0$. It suggests a traditional wisdom of economics: the law of demand. When ambulatory care becomes more expensive relative to inpatient care, patients will derive less utility from it. In other words, patients who receive additional financial assistance on hospitalization will be less likely to use primary care, holding other variables constant. Second, the magnitude of this marginal effect increases with $\theta$. Therefore, insurance policy affects severe patients more. This aligns with our empirical evidence that cost-sharing of hospitalization has heterogeneous effects on patients. (See Table \ref{tab:use/insptyl}.) We also find this marginal effect does not depend on weighting parameter $\gamma$. 
	
	\begin{figure}
		\includegraphics[width=0.48\textwidth]{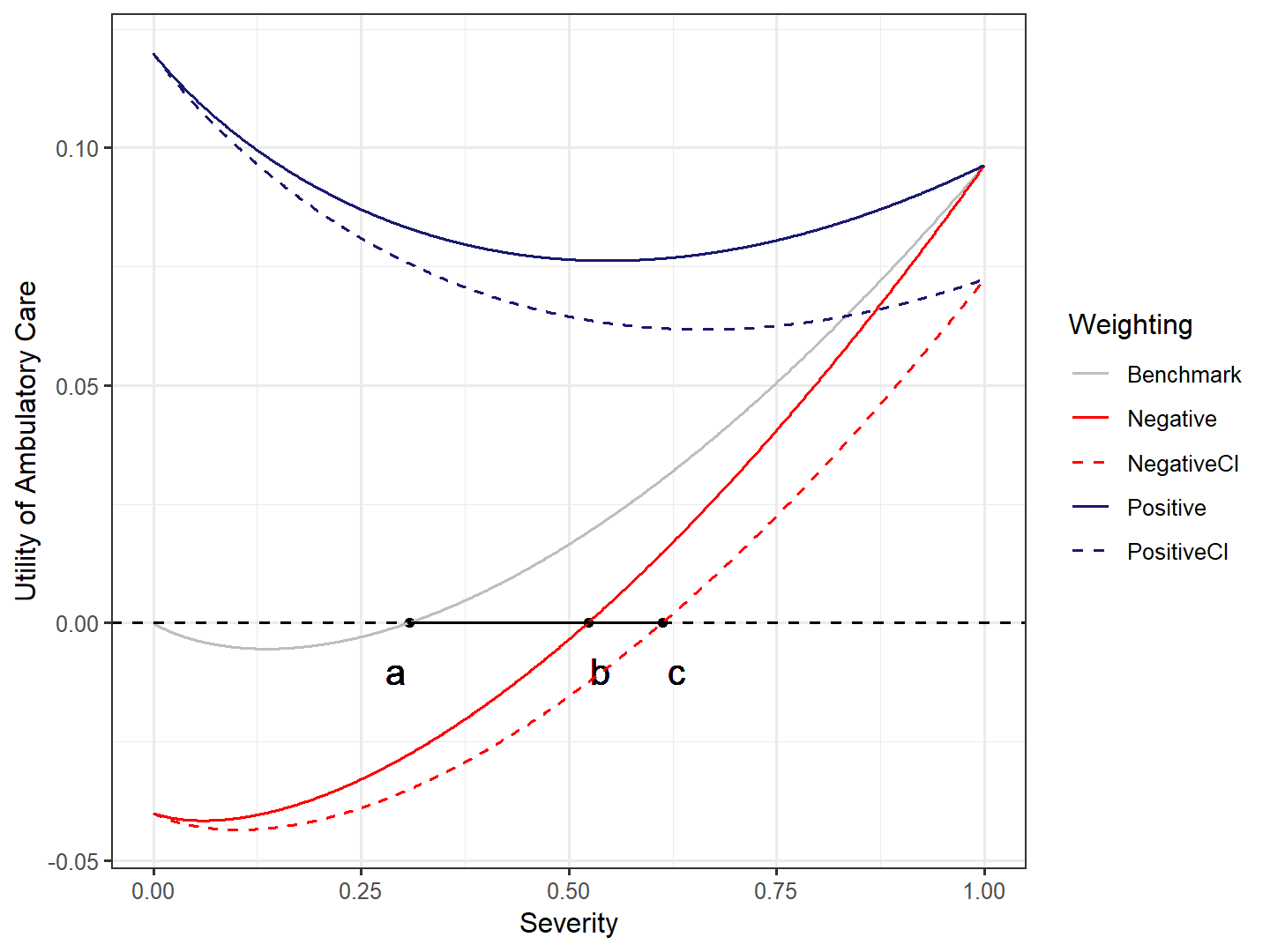}
		\caption{Utility of Using Ambulatory Care, effects of Insurance Policy $\frac{\phi_{pc} }{\phi_{hc}}$}
		\label{Fig:model/utci}
	\end{figure}
	
We then show the effect of cost-sharing ratio in Figure \ref{Fig:model/utci}. The parameters are specified as before. The solid curves represent the case where the cost-sharing ratio $\phi_{pc}/\phi_{hc}$ is one, the benchmark situation as in Figure \ref{Fig:model/utha}. We then increase the cost-sharing ratio to 1.2 and plot the utilities for each weighting parameter with the dashed curves. The differences between the solid and dashed curves show the effect of insurance policy for different weighting.
	
Generally speaking, the utility decreases in cost-sharing ratio, regardless of the weighting. Moreover, this effect of insurance policy is larger for patients with worse conditions. This finding is consistent with the evidence in the literature that the sickest are more likely to adjust healthcare utilization in response to changes in cost-sharing. (See \cite{BaickerJEP2011}.) 
	
As the insurance policy reduces the utility from using ambulatory care, the patients with negative weighting decide using primary care whenever $\theta_i > c$. Consequently, the insurance policy further induced the patients with health conditions in $[b,c]$ to abandon ambulatory care, thereby exacerbating the increase in total medical costs. 

Choosing an increased ratio of cost-sharing is motivated by the anti-poverty policy in our application. The new policy introduced the financial assistance to the disadvantaged population on their medical costs of hospitalization. It is reasonable at outset, because the intensive inpatient care delivers financial burden to treating catastrophic illness. Meanwhile, however, this insurance policy may unintentionally change the financial incentive to the disadvantaged for utilizing ambulatory care. As the ambulatory care becomes more expensive relative to hospitalization, patients shall use more inpatient care. As the simulation in our model suggests, the policy likely further drives up healthcare costs. We will bring this insight to data and examine whether it is indeed so in our application.

\section{Estimation}
	
We need to recover the following set of key model primitives from data: $\left\{ \alpha, \beta, \lambda,  \gamma, T\right\}$. 
We will estimate these parameters with two steps. First, $\alpha$ and $\beta$ are the parameters to measure how disease severity matters to medical costs; $\lambda$ is the effectiveness of ambulatory care. We will employ the information from data on healthcare costs to estimate these parameters. In particular, $\alpha$ and $\beta$ are identified from variations in medical costs of ambulatory care and inpatient care, respectively. The identification of $\lambda$ comes from a comparison in the inpatient care costs between whether patients using ambulatory care or not. 
	
Whereas $\gamma$ and $T$ are the parameters to determine the utilization of ambulatory care. In the second step, we therefore will use care utilization data to estimate how patient's weighting and opportunity costs determine using the ambulatory care.

\subsection{Explaining the costs of medical care} \label{paracost}
	
We take logarithm transformation of medical cost functions in section \ref{cost} to get: 
$$lnP_i^{pc} - ln\bar{P} = \alpha  ln\theta_i,$$
$$lnE(P_{i,0}^{hc}|P_{i,0}^{hc}>0) - ln\bar{K}^s = \beta ln\theta_i, $$
$$lnE(P_{i,1}^{hc}|P_{i,1}^{hc}>0) - ln\bar{K}^s = \beta ln\theta_i + \beta ln\lambda.$$
	
These equations offer a framework to estimate the model using the data at hand. We observe directly from data the ambulatory care costs, the hospitalization costs whether the patient has used ambulatory care or not. Meanwhile, we can also recover preference-discounted disease severity with the information from out-of-sample variation.  Details are provided in Appendix \ref{severity}. We shall translate these relations in the model to regression equations for estimation. 
	
First, we explain the ambulatory care costs with patient's disease severity by the following equation. 
\begin{equation}\label{a}
	lnP_{i,t}^{pc} = \alpha \cdot ln\theta_i + X_{i,t} + \mu_t + \varepsilon_{i,t}
\end{equation}
$lnP_{i,t}^{pc}$ is logarithm of medical cost of outpatient services on chronic disease management for patient $i$ in year $t$. We also add other observed patient's demographics like age and sex, and year fixed effect. Without $\bar P$ in the specification, its impact is implicitly included in the constant term. Consequently, without sub-index $i$, all the patients are assumed to refer to a same benchmark case of maximum primary care cost.  
	
Second, we use the following unified equation to explain the hospitalization costs regardless using ambulatory care or not.
\begin{equation}\label{bl}
	lnP_{i,t,s}^{hc} = \beta \cdot ln\theta_i + \rho \cdot d_i + X_{i,t} + \mu_t + \delta_s + \varepsilon_{i,t,s}.
\end{equation}
$lnP_{i,t,s}^{hc}$ is the medical cost of CVD inpatient services for patient $i$ discharged from hospital $s$ in year $t$. We also include $\delta_s$, the hospital fixed effect. $d_i$ is indicator variable for using ambulatory care. As the theory framework suggests, $\beta ln\lambda $ accounts for the difference between the patients using ambulatory care or not. In turn, such a difference is approximated by $\rho$. Thus, the effectiveness of ambulatory care can be derived by $ \lambda = exp( \rho / \beta)$. 
	
With regards to the worst possible situations in the specification, we need to take care of the maximum hospitalization cost $\bar{K}^s$ for each hospital type $s$. One possible way is to specify the max level for each $s$ from the observations in the data. However, it may have introduced much data noise from various hospital types. Certain hospitals may have limited data variation when we examine along different dimensions. We therefore include the fixed effect for hospital types $\delta_s$ in the regression. We categorize hospitals into four types: township health centers, local traditional medicine hospital, local general hospital and non-local tertiary hospitals. Respectively, we denote them as $\mathbf{s} = (s_1, ..., s_4)$. We impose further an identifying assumption as $\bar{K}^{s_j} = s_j\bar{K}^{s_1}$ for $j=2,3,4.$ Therefore, the hospital fixed effect $\delta_s$ shall capture the effects of $lns_j$ for $j=2,3,4$.

\begin{table}[htbp]\centering
	\scriptsize
\def\sym#1{\ifmmode^{#1}\else\(^{#1}\)\fi}
\caption{Parameter Estimates from Medical Costs \label{tab:est/abld}}
\begin{threeparttable}
\begin{tabular}{l*{2}{c}}
\hline\hline
                &\multicolumn{1}{c}{(1)}&\multicolumn{1}{c}{(2)}\\
                &\multicolumn{1}{c}{Ambulatory Care}&\multicolumn{1}{c}{Inpatient Care}\\
\hline
$ln(\theta)$    & 0.882\sym{***} ($\alpha$)&    1.489\sym{***} ($\beta$)\\
                & (0.311)   &  (0.440)         \\
                $d=1$        &    &      -0.253\sym{**} ($\rho$) \\
 &           &  (0.121)         \\               
TCM hospital   &  &     1.574\sym{***} ($lns_2$)\\
                &  &    (0.206)         \\
General hospital     &   &    2.285\sym{***} ($lns_3$)\\
                & &    (0.181)         \\
Non-local hospital         & &      3.231\sym{***} ($lns_4$)\\
                & &    (0.226)         \\     
\hline
Observations    &    317 &       255         \\
Adjusted \(R^{2}\)&  0.039 &     0.348         \\
\hline\hline
\end{tabular}
\begin{tablenotes}[flushleft]
\note Robust standard errors in parentheses. We aggregate the hospitalization records for each patient every year and at each hospital type. Township health centers ($lns_1$) is the benchmark.
\end{tablenotes}
\end{threeparttable}
\end{table}

The regression results of \ref{a} and \ref{bl} are reported in Table \ref{tab:est/abld}. We find the effects of disease severity on hospitalization cost is larger than that on ambulatory care, ie, $\beta > \alpha$. Using ambulatory care reduces the hospitalization cost by 25.3\%. Moreover, the better the hospital, the more costly of the hospitalization.

\subsection{Deciding for ambulatory care}
We follow the framework of discrete choice model to understand patient's decision on using ambulatory care. Specifically, a patient can get utility from using ambulatory care as:
$$u_i = v_i + \xi_i ,$$
where $\xi_i$ is an unobserved taste of using primary care for patient $i$ which is assumed to follow type-I extreme value distribution. 
	
The observed component $v_i$ follows the decision rule in our model ie equation (\ref{eq:pcdecisionwcpay}): 
\begin{equation} \label{vi}
	v_i = (1-\lambda^{\beta}) s_{ij}\theta_i^{\beta} + \gamma_i \frac{(1 - \theta_i)s_{ij}}{\phi_{hc}} - \frac{\phi_{pc}}{\phi_{hc}}  \theta_i^{\alpha}\frac{\bar{P}}{\bar{K}^{s_1}} - \frac{t_i}{\phi_{hc}}
\end{equation}
where $\bar{K}^{s_1} = \frac{\bar{K}^{s}}{s_j}$. $t_i =  \frac{T_i}{\bar{K}^{s_1}}$ measures the travel cost relative to the maximal medical cost for treating at township health centers. As patients' choice of hospital type may reveal their preference for care, we separate such an influence from $\gamma$ by adjusting the health benefits $(1 - \theta_i)$ of patient $i$ for any given hospital type of her choosing $s_{ij}$.
	
We use a binomial logit model, and the probability for patient $i$ using ambulatory care is:
$$\sigma_{i} = Prob (u_i>0) = \frac{exp[v_i]}{1 + exp[v_i]}. $$
We observe each patient's choice $d_i$, which is one if ambulatory care is used and $d_i=0$ otherwise. Thus,  the log-likelihood function is:
$$L = \sum_{i=1}^{N}\{d_i \times log\sigma_{i} + (1-d_i) \times log(1-\sigma_{i})\},$$
where $N$ is the total number of patients in the sample. We implement the estimation with a maximum likelihood method and report the estimates in Table \ref{tab:est/cost}. We also reported other parameters from application in the table. For example, the cost-sharing parameters $\phi_{pc}$ and $\phi_{hc}$, and the ratio of maximal medical cost $\frac{\bar{P}}{\bar{K}^{s_1}}$ are directly from the data. 
	
We exclude patients not having any other diagnoses than CVDs in the first step estimation, because they do not offer information to infer their preference-discounted disease severity. However, for estimating the model with a more complete picture, we decide to include these patients again. We use the data variation of disease severity with a discrete measure by Mild, Moderate and Severe as described in Table \ref{Tab:data/ss}. We specify ($\theta_{Mild}, \theta_{Moderate}, \theta_{Severe}$) by the 99-th percentile of their values within each category. 
	
\begin{table}[htbp]\centering
	\footnotesize
	\def\sym#1{\ifmmode^{#1}\else\(^{#1}\)\fi}
	\caption{Parameter Specification \label{tab:est/cost}}
	\begin{threeparttable}
		\begin{tabular}{l*{3}{c}}
			\hline\hline
			\multicolumn{2}{c}{Estimated Parameters} &\multicolumn{2}{c}{Observed Parameters}\\\cmidrule(lr){1-2}\cmidrule(lr){3-4}
			$\alpha$    &   0.882  & $\phi_{pc} $    &   0.35 \\
			$\beta$     &    1.489   & $\phi_{hc}: poor $    &   0.15 \\
			$\lambda$    &    0.844   & $\phi_{hc}: others$ & 0.41 \\
			$s_1$    &   1   &    $\theta_{Mild}$   &   0.1  \\
			$s_2$    &   4.816 &    $\theta_{Moderate}$    &   0.48\\
			$s_3$    &   9.836  & $\theta_{Severe}$    &   0.72\\
			$s_4$    &   25.103  &   $\bar{P}/\bar{K}^{s_1}$    &   0.7795  \\
			
			\hline\hline
		\end{tabular}
		\begin{tablenotes}[flushleft]\footnotesize
			\note ``Poor" refers to patients belong to the poor household identified for the welfare programs.
		\end{tablenotes}
	\end{threeparttable}
\end{table}
	
The key parameter for estimation is $\gamma$ which shows how much a patient prioritize disease prevention to monetary gain of using ambulatory care. Ideally, one can identify such a parameter for each individual. However, we do not have sufficient data for this purpose. Instead, we divide the population into different groups, within which the patients have same weighting in managing current condition from deteriorating. 
	
In this study, we lay our focus on the group of disadvantaged, whose household either resides at least 12 kilometers away from county center, or participates in various welfare programs because of its low annual income. Therefore, we estimate two groups with different weighting parameters in the benchmark case: $\gamma_h$ for regular households and $\gamma_l$ for the disadvantaged ones.
	
We include the travel cost in our model. This cost should be regarded as the opportunity cost of time for one to use ambulatory care. We further allow for heterogeneity in this travel cost across the population. We consider the opportunity value of the time may vary with the two dimensions along patient's demographics: income-level and sex. For example, the patients with high-income may have to incur more loss due to their absence from work to ambulatory care service. Similarly, male patients could be more critical manpower inputs for farm production. Therefore, we consider two additional parameters to measure travel cost for these patients, $t_H$ and $t_M$ for adjustments from benchmark $t_b$. The travel cost for high-income patients is $t_b + t_H$, and for male patients is $t_b+t_M$. 
	
Our identification strategy works as follows. Generally speaking, we take advantage of data variations in disease severity and in patient demographic characteristics to identify our structural parameters. In particular, within a benchmark sub-population with only low-income and female patients, their different choices in utilizing ambulatory care over varying disease severity deliver the identification of $\left\{ \gamma_l, t_b\right\}$. Then, we use other data variations in the dimensions of observed patient characteristics to pin down other parameters of interests. Specifically, patient's residence location, income-level and sex, help us to identify $\left\{ \gamma_h,  t_H, t_M \right\}$, respectively. 

\begin{table}[htbp]\centering
	\footnotesize
	\def\sym#1{\ifmmode^{#1}\else\(^{#1}\)\fi}
	\caption{Maximum Likelihood Estimates for Utility Parameters \label{tab:est/ut}}
	\begin{threeparttable}
		\begin{tabular}{l*{2}{c}}
			\hline\hline
			&\multicolumn{1}{c}{Coefficient}&\multicolumn{1}{c}{SE}\\
			\hline
			\emph{Panel A. Weighting}  & & \\
			$\gamma_h$: regular household     & 0.0225  & 0.0009 \\
			$\gamma_l$: disadvantaged         & -0.0166 & 0.0007 \\
			&&\\
			\emph{Panel B. Travel Costs}      & & \\
			$t_b$: benchmark                  &  0.1001 & 0.0067 \\
			$t_H$: adjustment for high-income &  0.4854 & 0.0092 \\
			$t_M$: adjustment for male        &  0.1166 & 0.0077 \\
			\hline\hline
		\end{tabular}
		\begin{tablenotes}[flushleft]\footnotesize
			\note Bootstrap standard errors are reported.
		\end{tablenotes}
	\end{threeparttable}
\end{table}
	
Our maximum likelihood estimates for the utility parameters are reported in Table \ref{tab:est/ut}. First, patients from regular households have a positive weighting parameter, indicating their concern for preventing their conditions from worsening through ambulatory care. On contrast, the disadvantaged patients have a negative parameter in managing their conditions. 
	
Recall that patients determine their utilization of ambulatory care based on the comparison between savings on expected healthcare costs and benefits from preventing deterioration,
$$\phi_{hc} (1-\lambda^{\beta}) s_{ij}\theta_i^{\beta} - \phi_{pc}\theta_i^{\alpha}\frac{\bar{P}}{\bar{K}^{s_1}} - t_i >-\gamma_i(1 - \theta_i)s_{ij}.$$
For regular patients with moderate conditions who opt for type 2 facilities for hospitalization, the perceived value of preventing their condition from deteriorating amounts to RMB 725 (equivalent to USD 110). In essence, these patients would choose ambulatory care even if the expected healthcare costs were to increase by RMB 725. Conversely, disadvantaged patients would refrain from utilizing ambulatory care until their healthcare costs could be reduced by RMB 535.
	
Second, the travel cost of using ambulatory care service is positive in general. On average, the travel cost for low-income female patients is about 10\% of the maximal hospitalization cost at township health clinics, which is RMB 630 (equivalent to US\$95). These opportunity costs become more significant if the patient is male, or has high-income. In particular, the high-income patients incur 48.5\% more in cost. 	Contrary to conventional assumptions that disadvantaged patients face greater challenges in accessing healthcare, our research suggests a different reality. Disadvantaged patients appear to encounter fewer opportunity costs when accessing ambulatory care services.

\subsubsection{Heterogeneity in Socioeconomic Status}
We have shown that the disadvantaged patients tend to have a negative parameter when managing their health conditions. We now turn to examine how patients with varying socioeconomic statuses may differ in their weighting of disease prevention versus financial savings. In this subsection, we specifically analyze the impact of two key factors.
	
First, a resident's household registration status, known as Hukou, may affect their weighting. Household registrations determine entitlement to government resources such as education, employment, and medical services. Nationwide, households are categorized as either urban or rural. In our analysis, we include a variable indicating the household's urban or rural registration status and introduce the parameter $\gamma_R$ to adjust for weighting parameter among rural residents.
	
It's important to note that all individuals in our dataset participated in the resident insurance scheme, ensuring equal access to medical services regardless of their urban or rural status. Additionally, these individuals are not employed in formal jobs, as indicated by their non-enrollment in the employee insurance scheme. Therefore, differences between the urban and rural residents in this application primarily stem from disparities in access to education and other public services.
	
Second, ethnic minority patients residing in villages associated with Tibetan tribes often share a common language, cultural practices, and unwritten societal norms that differ from those of the mainstream society. These minority patients are likely to adhere to more traditional social norms, which may impact their weighting. To account for this influence, we introduce the parameter $\gamma_M$ to adjust for weighting parameter among minority residents.
	
We then estimate in total of six groups with different weightings by four parameter: regular household registered in urban system ($\gamma_h$), majority disadvantaged household registered in urban system ($\gamma_l$), regular household registered in rural system ($\gamma_h +\gamma_R$), majority disadvantaged household registered in rural system ($\gamma_l +\gamma_R$), minority disadvantaged household registered in urban system ($\gamma_l +\gamma_M$), minority disadvantaged household registered in rural system ($\gamma_l +\gamma_R +\gamma_M$).\footnote{Here, we omit adjustments for minority among regular households because all minority residents live more than 12 kilometers away from the county center and thus fall under the disadvantaged category.} 

\begin{table}[htbp]\centering
	\footnotesize
	\def\sym#1{\ifmmode^{#1}\else\(^{#1}\)\fi}
	\caption{Parameters Estimates: Adjustment for Rural \& Minority \label{tab:est/utall}}
	\begin{threeparttable}
		\begin{tabular}{l*{2}{c}}
			\hline\hline
			&\multicolumn{1}{c}{Coefficient}&\multicolumn{1}{c}{SE}\\
			\hline
			\emph{Panel A. Weighting}  & & \\
			$\gamma_h$: regular household in urban      & 0.0211  & 0.0009 \\
			$\gamma_l$: majority disadvantaged in urban & -0.0134 & 0.0009 \\
			$\gamma_R$: adjustment for rural household  & -0.0016 & 0.0008 \\
			$\gamma_M$: adjustment for minority         & -0.0369 & 0.0025 \\
			&&\\
			\emph{Panel B. Travel Costs}      & & \\
			$t_b$: benchmark                  &  0.0989 & 0.0075 \\
			$t_H$: adjustment for high-income &  0.4512 & 0.0091 \\
			$t_M$: adjustment for male        &  0.1248 & 0.0077 \\
			\hline\hline
		\end{tabular}
		\begin{tablenotes}[flushleft]\footnotesize
			\note Bootstrap standard errors are reported.
		\end{tablenotes}
	\end{threeparttable}
\end{table}
	
We present our maximum likelihood estimates for utility parameters, adjusting for rural and minority status, in Table \ref{tab:est/utall}. Our analysis suggests a negative impact associated with rural registration and minority status. Specifically, rural residents, both from regular households and disadvantaged backgrounds, exhibit 7.6\% and 11.9\% lower weighting towards disease prevention, respectively, compared to their urban counterparts. However, among disadvantaged patients, the weighting parameter is 2.75 times higher for minorities than for those in majority groups. These findings suggest that social norms within ethnic minority communities have a greater influence on how patients prioritize disease prevention than their access to public resources, such as education.

\subsection{Robustness Check}
We switched to using the discrete measure on patient's severity when estimating the parameters for weighting and travel costs. It puts us in the position to face a challenge - whether those patients without any other diseases than CVDs should be considered as Mild. In order to address this concern, we next exclude these patients and repeat the estimation. We report the estimation results in Table \ref{tab:est/ut/rb1}.

\begin{table}[htbp]\centering
\footnotesize
\def\sym#1{\ifmmode^{#1}\else\(^{#1}\)\fi}
\caption{Parameter Estimates: Alternative Disease Severity Measures \label{tab:est/ut/rb1}}
\begin{threeparttable}
\begin{tabular}{l*{3}{c}}
\hline\hline
	&\multicolumn{1}{c}{(1)}&\multicolumn{1}{c}{(2)}&\multicolumn{1}{c}{(3)}\\
	&\multicolumn{1}{c}{Moderate/Severe}&\multicolumn{1}{c}{5 categories}&\multicolumn{1}{c}{Preference-discounted}\\
\hline
	\emph{Panel A. Weighting} & & &\\
	$\gamma_h$: regular household & 0.0031 &  0.0071 & 0.0094  \\
	$\gamma_l$: disadvantaged     & -0.0200 & -0.0162 & -0.0066 \\
	&&&\\
	\emph{Panel B. Travel Costs} & && \\
	$t_b$: benchmark                  & 0.1099 & 0.0912 & 0.1277 \\
	$t_H$: adjustment for high-income & 0.5187 & 0.4852 & 0.3867 \\
	$t_M$: adjustment for male        & 0.0565 & 0.05723 & 0.0408 \\
\hline\hline
\end{tabular}
\end{threeparttable}
\end{table}
	
Column (1) of Table \ref{tab:est/ut/rb1} shows the estimates from using the sample with only Moderate and Severe patients. We then further categorize patients with Moderate and Severe conditions into five bins by their inpatient cost on treating other diagnoses. The model estimates are shown in the column (2). In the end, we use the preference-discounted measure of disease severity constructed in appendix \ref{severity} to estimate the model. The result is in the column (3). In all specifications, our main findings remain qualitatively valid.

\section{Evaluation of Anti-poverty Programs}
The anti-poverty campaign in this county seeks to enhance the affordability of healthcare for the community. In this section, we examine the impact of the existing insurance policy on the utilization and cost of healthcare for disadvantaged patients.  Eventually, we would like to compare the effectiveness of alternative policy tools in offering assistance to the disadvantaged patients. 
	
To this end, we first compute the estimated utility of using ambulatory care $\hat{v}_i$ by substituting the estimated coefficients into equation (\ref{vi}). Thus, the predicted probability of using ambulatory care for each individual patient becomes $$\hat{\sigma}_i = \frac{exp[\hat{v}_i]}{1 + exp[\hat{v}_i]}.$$

Then, the predicted share of using ambulatory care among the designated patients under investigation is 
$$\hat{S}= \frac{\sum_{i=1}^{\tilde{N}}d_i \cdot \hat{\sigma}_i}{\sum_{i=1}^{\tilde{N}}[d_i \cdot \hat{\sigma}_i + (1-d_i) \cdot (1-\hat{\sigma}_i)]}$$ 
where $\tilde{N}$ is the number of patients in the designated group.
	
we compute the predicted healthcare cost (relative to the maximal hospitalization cost in township health centers) as follows,
\begin{equation*}
	\hat{P}_{i}^{total} = \hat{\sigma}_i \cdot [\hat{s}_{ij}(\hat{\lambda}\theta_i)^{\hat{\beta}} + \theta_i^{\hat{\alpha}}\frac{\bar{P}}{\bar{K}^{s_1}}] + (1-\hat{\sigma}_i) \cdot \hat{s}_{ij}\theta_i^{\hat{\beta}}
\end{equation*}

Moreover, the predicted patient welfare is computed by 
$$\hat{W}_i = \hat{\sigma}_i \cdot \hat{v}_i$$

\begin{table}[htbp]\centering
	\footnotesize
	\def\sym#1{\ifmmode^{#1}\else\(^{#1}\)\fi}
	\caption{Welfare Loss from the Current Insurance Policy for Disadvantaged Patients \label{tab:ct/loss}}
	\begin{threeparttable}
		\begin{tabular}{l*{3}{c}}
			\hline\hline
			&\multicolumn{1}{c}{With Assistance}&\multicolumn{1}{c}{W/O Assistance}&\multicolumn{1}{c}{Difference}\\
			&\multicolumn{1}{c}{(1)}&\multicolumn{1}{c}{(2)}&\multicolumn{1}{c}{(3)}\\
			\hline
			Ambulatory Care Use   &   0.145  & 0.189 & 0.044(23.13\%)  \\
			Expected Healthcare Cost     &   3.537  &  3.491 & -0.046(-1.33\%)  \\
			Patient Welfare        &  -0.090  &   -0.026  & 0.064(46.23\%) \\
			\hline\hline
		\end{tabular}
		\begin{tablenotes}[flushleft]\footnotesize
			\note Expected healthcare cost is relative to the maximum cost (around RMB 6300). Percentage changes are reported in the parentheses.  
		\end{tablenotes}
	\end{threeparttable}
\end{table}

In Table \ref{tab:ct/loss}, we present a comparison of predicted ambulatory care usage, total healthcare costs, and patient welfare for disadvantaged patients, with and without support from the current anti-poverty policy. Column (1) shows the baseline estimates from our model, which includes assistance. Column (2) reports the results under the scenario where medical assistance for the disadvantaged is removed, aligning their hospitalization cost-sharing with that of regular patients. Column (3) emphasizes the differences, illustrating the impact of removing the current anti-poverty policy on these disadvantaged individuals.	
	
Our findings reveal that under the current anti-poverty insurance policy, disadvantaged individuals are 23\% less inclined to utilize ambulatory care services. Consequently, this reluctance leads to an escalation in expected healthcare expenditures by RMB 290 (equivalent to USD 44) and results in an average welfare loss of 46.2\% per patient.

We next investigate two different policy tools targeting to improve the ambulatory care utilization by the disadvantaged patients. 
\begin{itemize}
	\item[Policy A.] Reducing cost-sharing for ambulatory care by 0.2 ($\downarrow$ Cost-sharing);
	\item[Policy B.] Offering lump-sum allowance for travel by RMB 200 (Travel Subsidy).
\end{itemize}
Reducing cost-sharing for ambulatory care by 0.2 is equivalent to 57.1\% change from current policy schemes. It averagely costs around 200 RMB per patient per year for those use ambulatory care. We introduce same amount as lump-sum subsidy to travel cost for each utilized patient per year. 

\begin{table}[htbp]\centering
	\footnotesize
	\def\sym#1{\ifmmode^{#1}\else\(^{#1}\)\fi}
	\caption{Policy Effects on Disadvantaged Patients \label{tab:ct/low}}
	\begin{threeparttable}
		\begin{tabular}{l*{5}{c}}
			\hline\hline
			&\multicolumn{1}{c}{}
			&\multicolumn{2}{c}{A. $\downarrow$ Cost-sharing}&\multicolumn{2}{c}{B. Travel Subsidy}
			\\\cmidrule(lr){3-4}\cmidrule(lr){5-6}
			&\multicolumn{1}{c}{(1)}&\multicolumn{1}{c}{(2)}&\multicolumn{1}{c}{(3)}&\multicolumn{1}{c}{(4)}&\multicolumn{1}{c}{(5)}\\
			&\multicolumn{1}{c}{Baseline}&\multicolumn{1}{c}{Counterfactual}&\multicolumn{1}{c}{Change}&\multicolumn{1}{c}{Counterfactual}&\multicolumn{1}{c}{Change}\\
			\hline
			Ambulatory Care Use   &   0.145  & 0.169 & 0.024(16.49\%)  & 0.156 & 0.010(6.88\%)\\
			Expected Healthcare Cost     &   3.537  &  3.505 & -0.032(-0.92\%) &  3.527 & -0.011(-0.30\%) \\
			Patient Welfare        &  -0.090  &   -0.036  & 0.053(38.77\%) &  -0.075  & 0.015(10.69\%)\\
			\hline\hline
		\end{tabular}
		\begin{tablenotes}[flushleft]\footnotesize
			\note Expected healthcare cost is relative to the maximum cost (around RMB 6300). Percentage changes are reported in the parentheses.  
		\end{tablenotes}
	\end{threeparttable}
\end{table}
	
We show the policy effects on the disadvantaged patients under experiments in Table \ref{tab:ct/low}.\footnote{We examine the policy effects on the regular patients under experiments as well. In general, the policies can also improve the ambulatory care utilization, expected healthcare cost, and patient's welfare for these regular patients as well.} First, both policies can effectively improve the ambulatory care utilization by the disadvantaged patients, and reduce the expected healthcare costs accordingly. Second, we assess their cost-effectiveness. To implement the new Policies A and B, the government expected to additionally spend RMB 32 and 30 per head, respectively. However, the total savings on the expected healthcare cost are RMB 202 and RMB 69, respectively. This means both policies are worth pursuing. Third, we measure patient's welfare in this analysis by her expected utility from using the care. Our results indicate that both policies can improve patients well being. 
	
In conclusion, reducing cost-sharing of ambulatory care is the most recommended policy. However, the existing welfare programs aim to offer affordable healthcare to the disadvantaged by lowering their hospitalization costs. Therefore, our work sheds light to policymakers for proper incentivizing the targeted population.

\section{Conclusion}
This paper analyzes insurance claims data from a rural county in China to explore the impact of anti-poverty programs on healthcare utilization and costs for disadvantaged patients. We show that disadvantaged individuals make different trade-offs between inpatient and ambulatory care when managing chronic conditions, compared to the general population. The current anti-poverty policy, which primarily reduces inpatient cost-sharing, inadvertently harms these patients by discouraging preventive care and raising overall healthcare costs. Lowering ambulatory care cost-sharing proves to be the most cost-effective strategy, as it reduces expenses and improves overall welfare for these individuals.
	
This work contributes to global efforts to fight against poverty and enhance healthcare equality by providing a deeper understanding of ambulatory care utilization among disadvantaged individuals. However, we acknowledge certain limitations in our study. First, our sample only includes individuals who have been hospitalized for cardiovascular diseases. As we cannot observe individuals at risk who have not yet utilized medical services, we are unable to estimate how ambulatory care affects disease prevention for a broader population. Investigating ambulatory care decisions for the general population remains a task for future research. Second, this work focuses on disadvantaged populations in rural, remote, and underdeveloped areas. We leave an effort to explore the utilization of ambulatory care with broader applicability to the future research.

\bibliographystyle{plain}
\bibliography{ref_primarycare.bib}
		
\newpage
\appendix
	
\section{Healthcare Delivery in the County} \label{map}
The sample county under study is a typical one in Western China. First, it is among the targets for China's anti-poverty campaign. The GDP per capita in 2018 was 31,100 RMB (equivalent to 4,690 USD), which was less than half of the national level. Second, it is geographically isolated and constrained. The county is located in a region with mountains, and its entire district is rather scattered. There are ten urban communities and 120 rural villages in this county, with more than 60\% of the general population being in rural areas. In addition, 34\% of the population is from an ethnic minority. They speak different languages or dialects, and tend to be more isolated from the county's center.

Third, the county has limited resources in term of offering medical services. Public hospitals in China are sorted into three grades by their treatment capacity: primary, secondary, and tertiary. Primary hospitals mostly function as clinics in a community by offering primary care and basic medical services. Secondary hospitals are not only larger in size, but also operate several clinical departments with specializations. The secondary hospitals are equipped with machines and resources that can offer diagnostic tests, X-rays, etc. Some basic surgeries such as appendectomies and cesarean sections are also available in these hospitals. However, neither of these hospitals can substitute for tertiary hospitals in terms of functionality and capability in China’s healthcare market.

There are 18 township health centers throughout the rural areas in the county. They function as local clinics and provide basic services in the community for common diseases such as cold. In addition, two secondary hospitals are available in the county center, with one serving only traditional Chinese medicine (TCM). Even though graded as secondary, the TCM hospital should be effectively regarded as a local clinic as it offers only basic medical services. 

The best medical care available in the county is offered by its General Hospital, as there is no tertiary hospital in this region. However, medical conditions with complication or challenges can only be treated in tertiary hospitals. Local residents have to travel by more than 5 hours of driving to the nearest tertiary hospital for more advanced medical services.

\begin{figure}
	\includegraphics[width=0.6\textwidth]{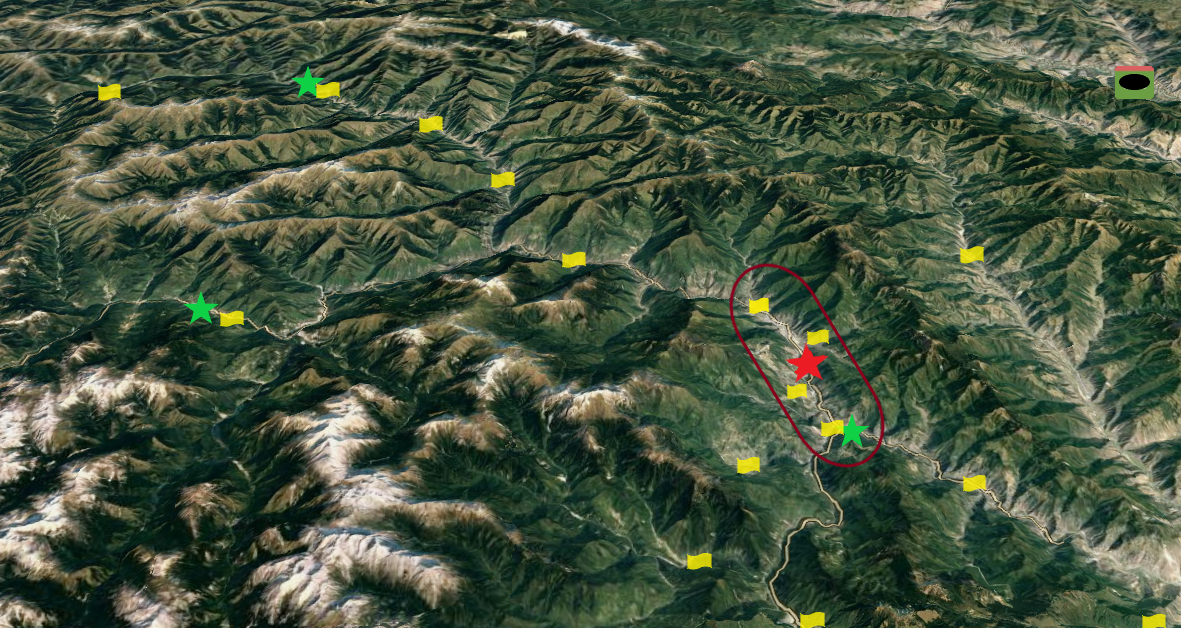}
	\caption{Map of the County}
	\label{Fig:map}
\end{figure}

Figure \ref{Fig:map} displays a map of the county. The red star represents the county center, which is also the location of the local general hospital. Yellow flags indicate the various townships within the county. Green stars denote township health centers, where ambulatory care for chronic disease management is provided.

This county features numerous mountains, resulting in challenging transportation and isolated villages. Patients residing in the farthest township (upper left yellow flag) must travel over 2 hours (approximately 90 kilometers) to reach the county center. However, this does not imply a lack of access to ambulatory care. The nearest ambulatory care provider (top green star) is situated around 30 km away, ensuring that these patients can still obtain the necessary healthcare services.

\begin{table}[htbp]\centering
	\footnotesize
\def\sym#1{\ifmmode^{#1}\else\(^{#1}\)\fi}
\caption{Share of Inpatients by Facility Type \label{tab:hos/pts}}
\begin{threeparttable}
\begin{tabular}{l*{4}{c}}
\hline\hline
			
                &\multicolumn{1}{c}{Cold}&\multicolumn{1}{c}{COPD}&\multicolumn{1}{c}{CVD}&\multicolumn{1}{c}{Cancer}\\
\hline
Type 1 Clinic &  51.30\% & 2.52\% & 17.29\% &  4.78\%   \\
Type 2 Hospital &  43.64\% & 90.72\% & 47.61\%  & 21.27\% \\
Type 3 Nonlocal &  5.05\% & 6.76\% & 35.10\%  & 73.95\% \\
\hline\hline
\end{tabular}
\end{threeparttable}
\end{table}

We classify treatment facilities in our sample in accordance with their actual treatment capacity. ``Type 1 Clinic" includes township health centers (THC) and the traditional Chinese medicine hospital (TCM). The local general hospital is considered as ``Type 2 Hospital", and the non-local tertiary hospitals are labeled as ``Type 3 Non-local". 

We first illustrate the status quo of medical services available in the county in Table \ref{tab:hos/pts}. For this purpose, we choose three representative and deadly chronic diseases, including CVDs (cardiovascular diseases), COPD (chronic obstructive pulmonary disease), and cancer, to show constrained treatment capacity demand attention and efforts to develop ambulatory care. 

Table \ref{tab:hos/pts} shows the shares of patient visits to different facilities by diagnosis. The county has limited medical resources in offering care. Its general hospital can at best handle treatment needs for COPD. Medical conditions involve further treatment capacity like CVDs and cancer are mostly treated at the tertiary hospitals outside the region. It suggests that local treatment facilities may be inadequate in handling severe acute cardiovascular events, such as stroke, heart attack, and heart failure. Since these situations can have disastrous consequences without prompt medical intervention, it is crucial to enhance ambulatory care in underdeveloped areas to prevent such acute events from occurring.

\section{Preference-discounted Measure of Disease Severity}
\label{severity}
	
We use out-of-sample variation of patient's hospitalization spending to infer her disease severity. In particular, we use the patient's inpatient care cost for treatment unrelated to CVDs to construct a continuous variable for her disease severity in our empirical analysis. 
	
We first use a regression to take out systematic impact by patients' observed characteristics on hospitalization spending. Specifically, we regress patient's hospitalization cost on her observed demographics, the fixed effects for treatment facility and year.
$$ln \tilde{P}_{i,t,s}^{hc} = a_1 + a_2 X_{i,t} + \mu_t + \delta_s +  \epsilon_{i,t,s}$$
where $\tilde{P}_{i,t,s}^{hc}$ is the non-CVD inpatient care cost for patient $i$, discharged from hospital $s$ in year $t$. We only select diagnoses with no provision of ambulatory care: respiratory, gastroenterology and orthopedics diseases, end-stage renal disease, and some surgical diseases, such as appendicitis, anorectal disease and trauma. 
	
The estimated residual $ \hat{\epsilon}_{i,t,s}$  is thus the unobserved factors correlated with hospitalization cost for non-CVDs. Our approach builds on the assumption that disease severity and patient preference for care should not vary over diagnoses. If other diseases had cost more for a patient, then her CVD conditions may likely to incur more cost than other CVD patients as well, holding other factors constant. 
	
We further summarize the patient's disease severity by calculating her weighted average residual $\bar \epsilon _i$ over hospitalization $h$, diagnosis $d$ and year $t$.
$$\bar{\epsilon}_i = \frac{1}{T\times D\times H} \sum_{t}\sum_{d}\sum_{h} \omega_d  \hat{\epsilon}_{i,t,s},$$
where $\omega_d$ is the weight by the frequency of diagnosis $d$. Finally, we standardize $\bar \epsilon$ so that it ranges from 0 to 1. Thus, the patient's preference-discounted disease severity is 
$$\theta_i = \frac{\bar{\epsilon_i} - \min\left\{\bar{\epsilon }_j : \forall  j\right\}}{ \max\left\{\bar{\epsilon }_j : \forall  j\right\}  - \min\left\{\bar{\epsilon }_j : \forall  j\right\}}.$$

\section{Model Prediction with Alternative Explanations}	\label{alternative}
\subsubsection{Time Preference}
When patients exhibit a preference for immediate rewards or heavily discount future benefits, they may under-utilize ambulatory care. This occurs because the benefits of ambulatory care, such as reducing hospitalization costs and preventing disease deterioration, are typically realized in the distant future, whereas the immediate costs associated with its utilization are apparent. For disadvantaged patients who significantly discount these delayed benefits, the immediate costs outweigh the long-term advantages, leading to reduced use of ambulatory care and an overall increase in healthcare expenditures.

We follow the canonical model of present bias (\cite{LaibsonQJE1997, OAER1999}) and introduce the preference for immediate versus delayed outcomes, denoted as $\delta_i$.\footnote{In addition to the present bias parameters, there is also time-consistent discount rate in their models. However, we set the time-consistent discount rate to 1 in our framework for simplification. } Then the utility from using ambulatory care when patients exhibit time preference can be specified as
\begin{equation} \label{discount}
	U_i^{\delta} = \delta_i \cdot [(1-\lambda^{\beta}) \theta_i^{\beta} \bar{K}^s + \gamma_i (1 - \theta_i)] - (\theta_i^{\alpha}\bar{P} + T_i).
\end{equation}
It's worth noting that, under an alternative specification, a discount factor concerns only future hospitalization costs. (ie, $\delta_i (1-\lambda^{\beta}) \theta_i^{\beta} \bar{K}^s $) In a case $\delta$ is close to 0, this alternative specification can account for situations where patients may be inattentive to hospitalization costs (\cite{AbaluckAER2011}). These specifications provide similar insights regarding the model's implications, so we will discuss only specification \ref{discount} here.

\begin{figure}
	\includegraphics[width=0.48\textwidth]{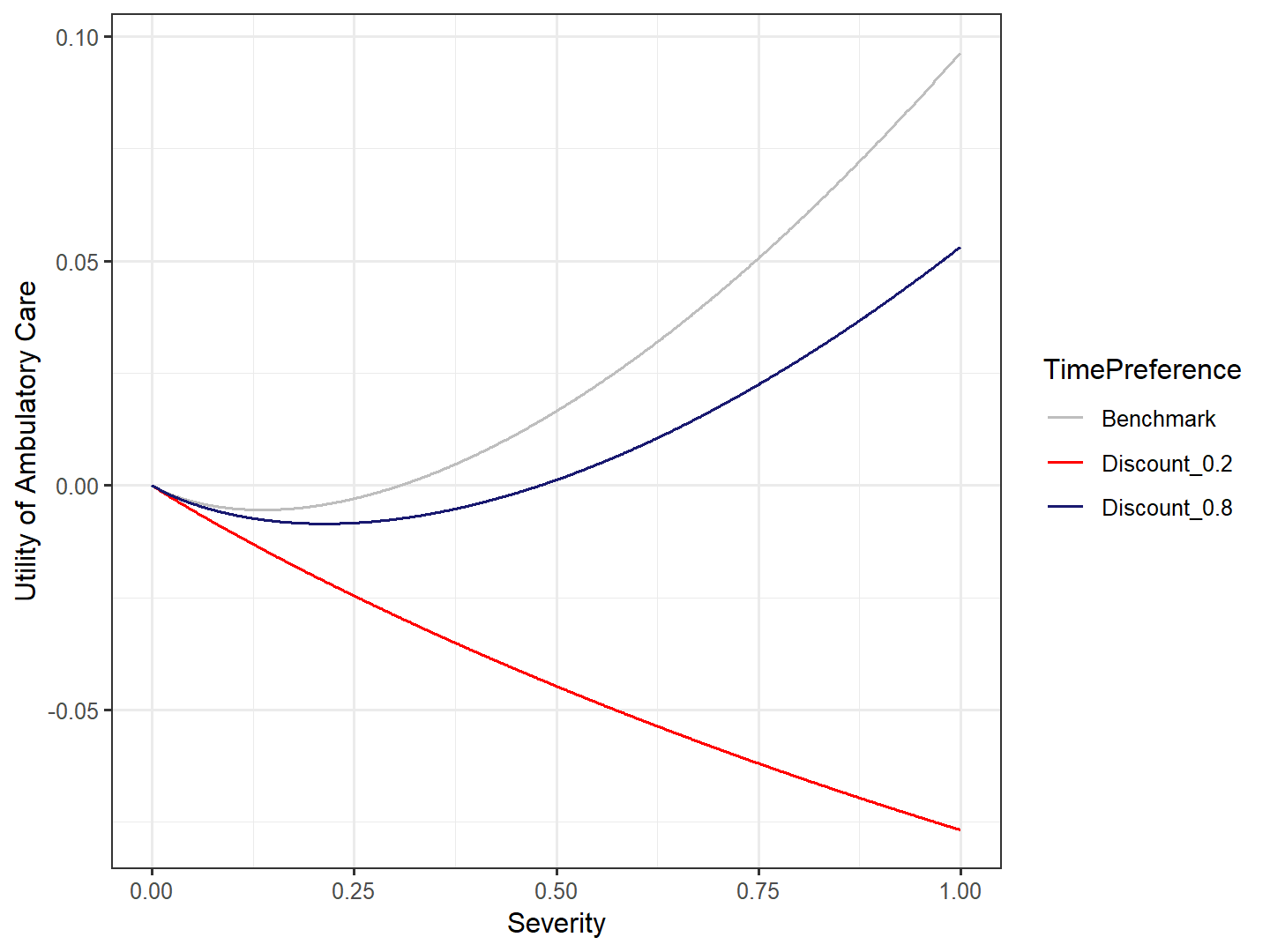}
	\caption{Utility of Using Ambulatory Care (by Time Preference $\delta$)}
	\label{Fig:model/uthdc}
\end{figure}

We then plot the utility of using ambulatory care over disease severity for different time preference in Figure \ref{Fig:model/uthdc}. The gray curve represents the baseline case with $\gamma=0$ and $\delta=1$, indicating simple monetary trade-off. The blue curve corresponds to a smaller discount factor ($\delta=0.8$), while the red curve represents a larger discount factor ($\delta=0.2$). We isolate the effect of time preference by setting $\gamma=0$. The other parameters are specified as before. 

In our model, medical costs are assumed to rise with disease severity. Therefore, a greater present bias leads to higher utility from using ambulatory care for the patients with less severe conditions, because their immediate costs are lower. Conversely, patients with a lesser present bias tend to wait until their condition worsens, when the delayed benefits become more pronounced. As a result, variations in time preference contribute to differences in ambulatory care utilization, particularly among those with more severe conditions. This pattern is rather different from what we saw in the data  (Figure \ref{Fig:use/pred1}). Therefore, using variation in time preference alone cannot find a way to interpret the ambulatory care choice in the disadvantaged.

\subsubsection{Symptom Salience}
Patients often overlook less salience symptoms, leading to under-utilization of treatments designed to address non-painful symptoms. Since many initial symptoms of chronic conditions, like high glucose or blood pressure levels, may not be readily apparent, patients might not recognize these conditions and consequently underestimate the benefits of seeking ambulatory care for managing their health. 

We adopt the model proposed by \cite{BaickerQJE2015}, which in turn is developed from the limited attention framework in \cite{DellavignaJEL2009}. We assume that the severity of chronic conditions, $\theta$, is solely determined by painful symptoms, $n$. That is, $\theta = n.$ Subsequently, inattentive patients tend to underestimate symptoms and act on a ``decision severity'', instead of the true disease severity $\theta$, 
$$\hat{\theta} = \mu n = \mu\theta, $$ 
where $0 <\mu \leq 1$. Therefore, the utility from using ambulatory care in cases where patients exhibit inattention to non-salient symptoms can be specified as
\begin{equation} \label{salience}
	U_i^{\mu} = (1-\lambda^{\beta}) (\mu_i\theta_i)^{\beta} \bar{K}^s + \gamma_i (1 - \mu_i\theta_i) - [(\mu_i\theta_i)^{\alpha}\bar{P} + T_i]
\end{equation}

\begin{figure}
	\includegraphics[width=0.48\textwidth]{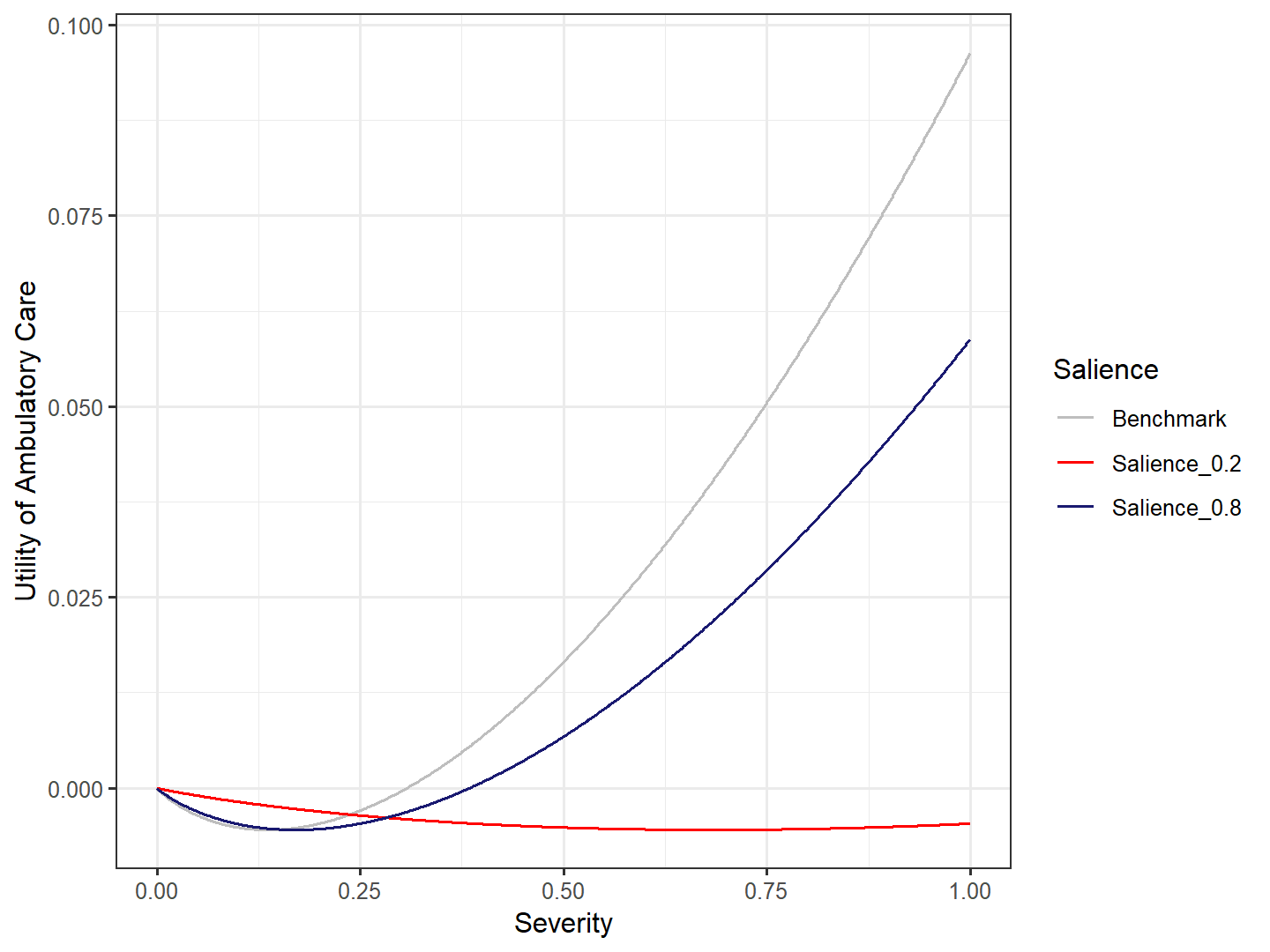}
	\caption{Utility of Using Ambulatory Care (by Salience $\mu$)}
	\label{Fig:model/uths}
\end{figure}

We then plot the utility of using ambulatory care over disease severity for different degree of inattention in Figure \ref{Fig:model/uths}. The gray curve represents the baseline case with $\gamma=0$ and $\mu=1$. The blue curve corresponds to less inattention case ($\mu=0.8$), while the red curve represents larger inattention case ($\mu=0.2$). Again, we isolate the effect of salience by setting $\gamma=0$. The other parameters are specified as before. 

Variation in inattention to non-salient symptoms result in different utilization in ambulatory care among more severe patients. This is so because the inattention makes more severe patients over-confident with their conditions. Such miscalculation leads to underestimate the benefits of ambulatory care.  As a result, this factor of inattention to non-salient symptoms cannot explain the observed differences in ambulatory care utilization between disadvantaged patients and others.

\subsubsection{Biased Beliefs}
Biased beliefs can influence patients' perceptions of treatment effectiveness, which in turn serve as a potential explanation for under-utilization. In order to model this factor, we introduce biased beliefs, $\tilde{\lambda}$, which is different from the true effectiveness of ambulatory care, $\lambda$. For example, when $\tilde{\lambda} < \lambda$, patients overestimate the effectiveness of ambulatory care. The utility from using ambulatory care can then be expressed as:
\begin{equation} \label{beliefs}
	U_i^{\lambda} =(1-\tilde{\lambda}_i^{\beta}) \theta_i^{\beta} \bar{K}^s + \gamma_i (1 - \theta_i) - (\theta_i^{\alpha}\bar{P} + T_i)
\end{equation}

\begin{figure}
	\includegraphics[width=0.48\textwidth]{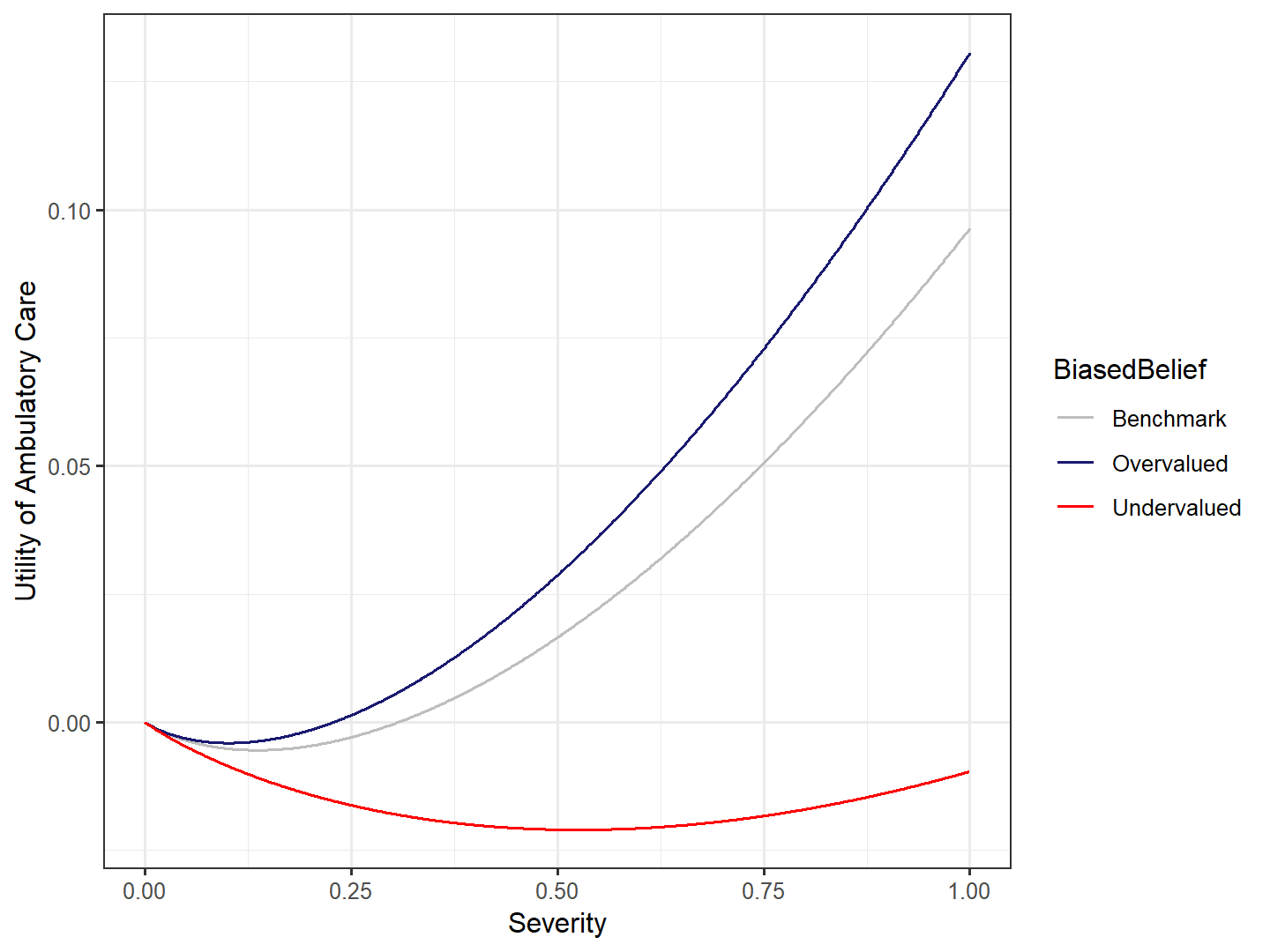}
	\caption{Utility of Using Ambulatory Care (by Belief $\tilde{\lambda}$)}
	\label{Fig:model/uthbf}
\end{figure}

We plot the utility of using ambulatory care over disease severity for different types of biased beliefs in Figure \ref{Fig:model/uthbf}. The gray curve represents the baseline case with $\gamma=0$ and $\tilde{\lambda}=\lambda=0.85$. The blue curve corresponds to overvalued case ($\tilde{\lambda}=0.825$), while the red curve represents undervalued case ($\tilde{\lambda}=0.925$). Again, we isolate the effect of biased beliefs by setting $\gamma=0$. The other parameters are specified as before. 

In fact, our conclusions align with those of previous sections that examined other potential factors influencing the use of ambulatory care. The variations in biased beliefs alone cannot serve as the explanation for the observed differences in ambulatory care utilization among disadvantaged patients.

\end{document}